\documentclass[aps,twocolumn,a4paper,showkeys,footinbib,longbibliography]{revtex4-1}

\usepackage{amsmath,amssymb,bm,amsthm}
\usepackage{subfigure}
\usepackage{graphicx}
\usepackage{dcolumn}
\usepackage{bm}
\usepackage[mathlines]{lineno}
\usepackage{color}
\newcounter{one}
\usepackage{url}

\usepackage{textcase}

\usepackage{colortbl}
\usepackage{tabularx}
\usepackage{verbatim}
\usepackage{multirow}

\usepackage[T1]{fontenc} 
\usepackage{lmodern}
\usepackage{bbm}
\usepackage[utf8]{inputenc}
\usepackage{amsfonts}
\usepackage{array}

\usepackage{bbm}
\usepackage{enumitem}
\usepackage{umoline}
\usepackage[usenames,svgnames]{xcolor}
\usepackage{natbib}
\usepackage[hyperindex,breaklinks]{hyperref}
\hypersetup{
     colorlinks=true,       		
     linkcolor=Navy,          	
     citecolor=Navy,            
     filecolor=Navy,      		
     urlcolor=Navy,           	
    runcolor=cyan,
 }
\usepackage{graphicx}
\usepackage{amsfonts}
\usepackage{amssymb}
\usepackage{amsmath}
\usepackage{bbm}
\usepackage{enumitem}

\newcommand{\bra}[1]{\langle #1 |}
\newcommand{\ket}[1]{| #1 \rangle}

\newcommand{\tr}[0]{ {\rm tr}}

\newcommand{\half}[1]{{ \rm h}}
\newcommand{\Oorderof}{\mathcal{O}}
\newcommand{\orderof}[1]{\Oorderof(#1)} 

\newcommand{\for}[0]{\quad \textrm{for} \quad}

\newcommand{\dist}{{\rm dist}}

\newcommand{\poly}{{\rm poly}}

\def\beq{\begin{equation}}
\def\eeq{\end{equation}}
\def\nbeq{\begin{equation*}}
\def\neeq{\end{equation*}}
\def\<{\langle}
\def\>{\rangle}

\def\tr{{\rm tr}}

\newtheorem{theorem}{Theorem}
\newtheorem{lemma}{Lemma}

\newtheorem{definition}{Definition}  
\newtheorem{prop}[theorem]{Proposition} 

\newcommand{\sectionprl}[1]{{\par\it #1.---}}

\usepackage{ulem}

\begin{document}
\title{Eigenstate thermalization from the clustering property of correlation}

\author{Tomotaka Kuwahara$^{1,2}$ and Keiji Saito$^{3}$}
\affiliation{$^{1}$
Mathematical Science Team, RIKEN Center for Advanced Intelligence Project (AIP),1-4-1 Nihonbashi, Chuo-ku, Tokyo 103-0027, Japan}
\affiliation{$^{2}$Interdisciplinary Theoretical \& Mathematical Sciences Program (iTHEMS) RIKEN 2-1, Hirosawa, Wako, Saitama 351-0198, Japan}

\affiliation{$^{3}$Department of Physics, Keio University, Yokohama 223-8522, Japan}

\begin{abstract}
  The clustering property of an equilibrium bipartite correlation is one of the most general thermodynamic properties in non-critical many-body quantum systems. Herein, we consider the thermalization properties of a system class exhibiting the clustering property. We investigate two regimes, namely, regimes of high and low density of states corresponding to high and low energy regimes, respectively. We show that the clustering property is connected to several properties on the eigenstate thermalization through the density of states. Remarkably, the eigenstate thermalization is obtained in the low-energy regime with sparse density of states, which is typically seen in gapped systems. For the high-energy regime, we demonstrate the ensemble equivalence between microcanonical and canonical ensembles even for subexponentially small energy shell with respect to the system size, which eventually leads to the weak version of eigenstate thermalization.

\end{abstract}
\maketitle

\sectionprl{Introduction}
Thermalization in an isolated quantum system is a fundamental phenomenon that is directly connected to the arrow of time in nature. The first study on this phenomenon dates back to Von Neumann's study in 1929~\cite{vonNeumann}. Recently, this subject has been revived, fueled by relevant experiments~\cite{kinoshita2006quantum,trotzky2012probing,Gring1318,Kaufman794,PhysRevX.8.021030,PhysRevLett.121.023601,PhysRevLett.120.197701} and a new concept resulting from the quantum information theory~\cite{popescu2006entanglement,Goold_2016}. The studies have now become interdisciplinary, including statistical physics, quantum information theory, and experiments~\cite{Yukalov_2011,doi:10.1146/annurev-conmatphys-031214-014726,doi:10.1080/00018732.2016.1198134,Gogolin_2016,Mori_2018}.

One of the central subjects in thermalization is the eigenstate thermalization hypothesis (ETH) that guarantees the thermodynamic property of an isolated quantum system~\cite{Reimann_2015,Deutsch_2018}. The ETH states that an expectation value of any local observable for only one eigenstate is identical to the quantity calculated by the canonical ensemble with the corresponding inverse temperature~\cite{deutsch1991quantum,srednicki1994chaos}. So far, the ETH has been intensively studied using numerical calculations~\cite{kim2014testing,mondaini2016eigenstate,mondaini2017eigenstate,doi:10.1080/00018732.2016.1198134}, as well as by the phenomenological analyses based on random matrices \cite{Reimann_2015} for the regime of high density of states. Although various counterexamples exist~\cite{doi:10.1146/annurev-conmatphys-031214-014726, PhysRevLett.119.030601, Turner2018, PhysRevLett.122.173401, PhysRevB.98.235156}, nonintegrable systems are generally believed to satisfy the ETH in high-energy regimes. However, a theoretical rationale of the ETH in low-energy regime is still missing. Even with the help of numerical studies, it is also generally difficult to establish any conclusive results on ETH in the low-energy regime.

\begin{figure}[tt]
\centering
{
\includegraphics[clip, scale=0.4]{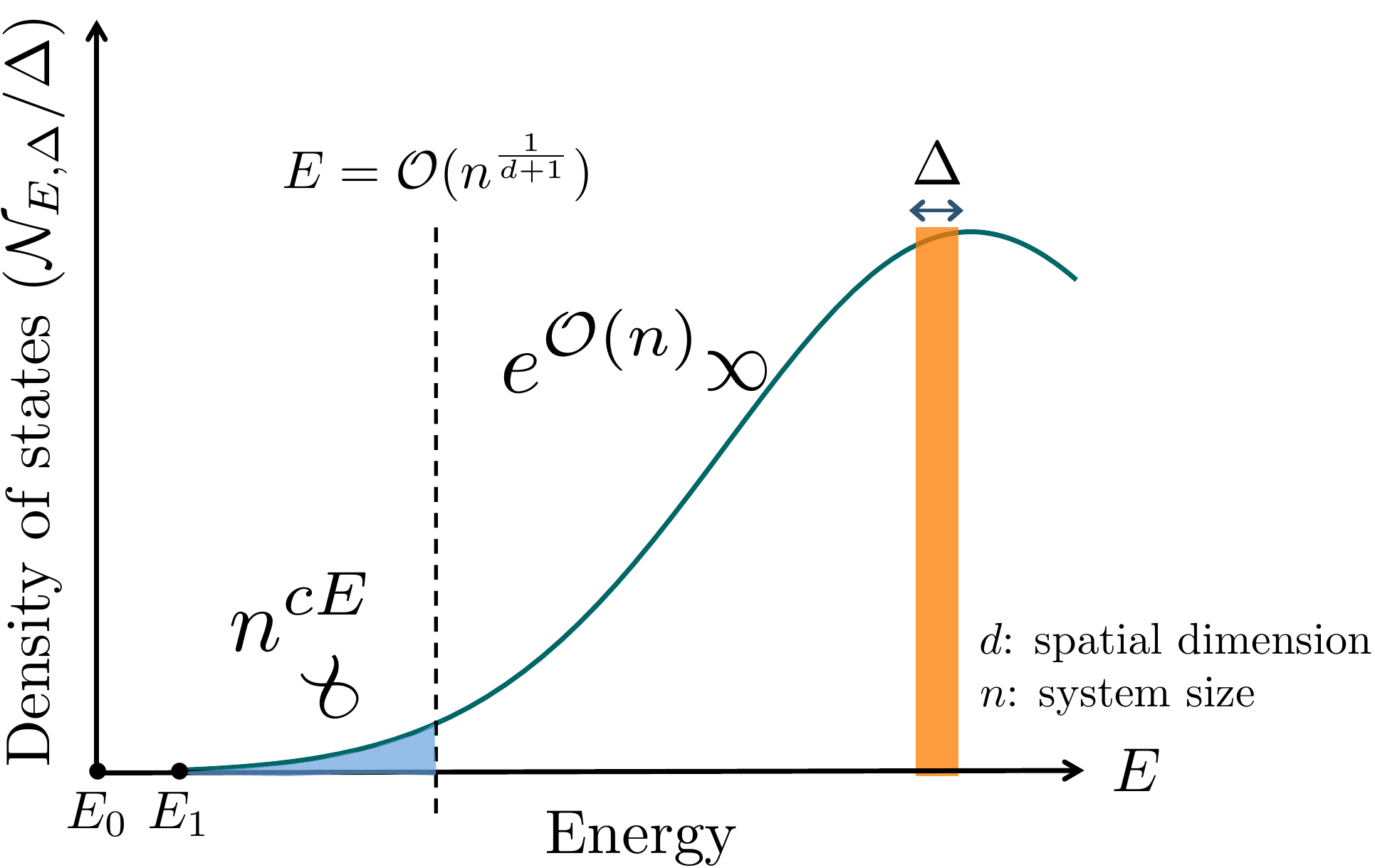}
}
\caption{(color online). 
We suppose a scaling of $n^{\orderof{E}}$ for the density of states, which typically appears in non-critical (or gapped) systems at low temperatures.
Subsequently, under the assumption of clustering (Def.~\ref{def:clustering}) of a canonical state, all the low-lying eigenstates are proved to demonstrate the same properties as those of the ground state (blue shaded). In high-energy regime, the weak version of the ETH and the ensemble equivalence between the microcanonical and canonical distributions hold as long as an energy width of $\exp[-\orderof{n^{\frac{1}{d+1}}}]$ is assumed.}
\label{result_outline}
\end{figure}

In this work, we discuss the eigenstate thermalization in the low-energy regime as well as the high-energy regime, focusing on the system class that satisfies {\it the clustering property} (i.e., exponential decay of bi-partite correlations, see Def.~\ref{def:clustering} below). The clustering property is one of the most general thermodynamic properties in noncritical systems, valid irrespective of characteristics such as the nonintegrability, energy regime, and so on. In this paper, a special attention is paid on the difference in the density of states between the low and high energy regimes, as the density of states can vary with respect to the energy regime~(See Fig.~\ref{result_outline}). We show that the clustering property is deeply connected to several properties on the eigenstate thermalization through the density of states.

To study the low-energy regime, we consider a system having density of states with a power-law dependence on the system size, typically observed in gapped systems \cite{PhysRevB.76.035114, PhysRevB.92.115134,PhysRevA.80.052104}. Note that the gapped systems satisfy the clustering property even for the ground state. Then, we report that the eigenstate thermalization is proven for low-lying energy eigenstates near the ground state. This implies that low-lying eigenstates behave similar to the ground state in the sense that these yield the same expectation values for local observables. This can also be considered as {\it zero-temperature version of eigenstate thermalization}. In the high-energy regime, we discuss the ensemble equivalence between microcanonical and canonical ensemble. We show that it holds for sub-exponentially small energy width with respect to the system size, i.e., $\Delta_c \sim \exp[-\orderof{n^{\frac{1}{d+1}}}]$, thus achieving an exponential improvement compared to the recent results in \cite{brandao2015equivalence,Tasaki2018}. Although this small energy width cannot reach the ETH since the energy spacing is even smaller, we quantitatively discuss the weak version of the eigenstate thermalization. It implies that the variance of the deviation from the ETH decays as the system size increases. We show that the decaying rate depends on $(\log n )^{d+1}/n$, where $n$ and $d$ are the system size and the spatial dimension, respectively. This is consistent with the recent numerical observations~\cite{PhysRevB.91.155123}.

\sectionprl{Setup of Hamiltonian}
For simplicity, we consider a (1/2)-quantum spin system defined on a $d$-dimensional hypercubic lattice. The present analysis is applied to generic quantum spin numbers and lattice structures. The Hamiltonian comprises $n$ local terms: 
\begin{align}
H= \sum_{i=1}^n H_i   , \quad \|H_i\| \le g
\label{eq:D_ham}
\end{align}
where $\|\cdot\|$ represents the operator norm. By taking the energy unit appropriately, we set $g=1$. The local Hamiltonian, $H_i$, contains spin operators that act on spins, $j$, with distance, $\dist(i,j)\le \ell_0$, where $\dist(i,j)$ is the Manhattan distance. Notably, translation invariance of the Hamiltonian is not assumed here. Subsequently, without loss of generality, we set the energy of the ground state to zero. In addition, we assume that the system satisfies the clustering property for the canonical distribution that is defined as follows:
\begin{definition} \label{def:clustering}
Let $\rho_{\beta}$ be a canonical distribution with an inverse temperature, $\beta$,  
\begin{align}
\rho_{\beta} & := e^{-\beta H}/ Z_{\beta} ,
\end{align}
where $Z_{\beta}$ is the partition function.
Let $A_X, B_Y$ be the arbitrary operators supported on subsets $X$ and $Y$, respectively. A density matrix $\rho_{\beta}$  is assumed to satisfy the $(r, \xi)$-clustering if the following condition is satisfied,
\begin{align}
\left| \langle A_X B_Y \rangle_{\beta} - \langle  A_X \rangle_{\beta} \langle B_Y \rangle_{\beta}\right| \le \| A_X \| \| B_Y \| e^{-\dist(X,Y)/\xi} \notag 
\end{align}
for $\dist(X,Y)\ge r$ with fixed constants, $r$ and $\xi$. Here, $\dist (X,Y)=\min_{i\in X , j \in Y} \dist (i,j)$ and $\langle ...\rangle_{\beta}=\tr (...\rho_{\beta})$.
\end{definition}
So far, several previous rigorous studies have addressed the clustering property for low and high energy regimes~\cite{Araki1969,1367-2630-17-8-085007,Park1995,ueltschi2004cluster,PhysRevX.4.031019,frohlich2015some,kuwahara2019Markov,Thomas1983,Kennedy1985,10.1143/PTPS.87.233,PhysRevB.56.5535}. 
The clustering property is thought to be a rather generic phenomenon in non-critical many-body quantum systems.
Note that no direct connection exists between the clustering property and the nonintegrability of the system; hence, even integrable systems can exhibit clustering. 

\sectionprl{Definition of the eigenstate thermalization}
We consider the macroscopic observable, $\Omega$, that is composed solely of local operators:
\begin{align}
    \Omega &= \sum_{i=1}^n \Omega_i  ,\quad \|\Omega_i\| \le 1,
\label{eq:op_def}
\end{align}
where $\Omega_i$ is composed of spin operators that act on spins, $j$, with  distance, $\dist(i,j)\le \ell$.
Throughout this paper, we set $\ell \ge r$. Using these observables, we define the eigenstate thermalization irrespective of the energy regime. 
Let $\ket{E_m}$ and $E_m$ be the $m$th eigenstate and the eigenenergy of the system, respectively ($m=0,\ldots,2^n-1$). For the macroscopic observable, the eigenstate thermalization is defined as
\begin{align}
|  \langle E_m | (\Omega /n)  | E_m \rangle - \langle \Omega/n  \rangle_{\beta} | \to 0, \quad n\to \infty  \, ,  \label{ee}
\end{align}
for all energy eigenstates within a given finite energy shell. The canonical average is computed with the temperature corresponding to the energy. The eigenstate thermalization here is defined as the ensemble equivalence between the single eigenstate and the canonical ensemble in terms of the given observables~\cite{Supplement_ETH_clustering}.

In case that one can not access the eigenstate thermalization described above, we consider the weak version of eigenstate thermalization to quantify ETH-like properties for general classes of systems including integrable systems~\cite{PhysRevLett.105.250401, PhysRevLett.119.100601}. To this end, we define the microcanonical distribution in the energy shell with the energy width $\Delta$:
\begin{align}
    \rho_{{E},\Delta} & := {\cal N}_{{E},\Delta}^{-1} \sum_{E_m \in ({E}-\Delta,{E}] } \ket{E_m}\bra{E_m} ,
    \label{Def:micro_cano_ensemble} 
\end{align}
where ${\cal N}_{{E},\Delta}$ is the number of eigenstates within the energy shell. Hence, ${\cal N}_{{E}, \Delta}/ \Delta$ is the density of states depicted in Fig.~\ref{result_outline}. Then, the weak version of eigenstate thermalization is written as 
\begin{align}
{\rm Var}_{{E},\Delta}(\Omega/n)  & \to 0 \, , ~~~~~~~~ n\to \infty \, , \label{wee}  \\
{\rm Var}_{{E},\Delta}(O) & :=\!\!\!\!\!\!\!\!\sum_{ E_m \in ({E}-\Delta,{E}]} \!\!\!\!\! \frac{ \left( \bra{E_m} O \ket{E_m} -  \langle O \rangle_{{E},\Delta}  \right)^2}{{\cal N}_{{E},\Delta}}  ,  \label{wee_var}
\end{align}
where $\langle O  \rangle_{{E},\Delta}$ is the microcanonical average for an arbitrary operator $O$.


\sectionprl{Results for low-energy regime}
We derive the following theorem, by which we can prove the eigenstate thermalization in the low energy regime if the density of states is sparse, as depicted in Fig.~\ref{result_outline}. The clustering property leads to the Chernoff-Hoeffding-type concentration inequality for macroscopic observables ~\cite{Kuwahara_2016,Anshu_2016}, which is crucial in the derivation. We provide the details of the derivation in the supplementary material~\cite{Supplement_ETH_clustering}.
\begin{theorem} \label{main_theorem_Ensemble Equivalence_Strong}  
  Let $\beta^\ast$ be an arbitrary inverse temperature for which the canonical ensemble, $\rho_{\beta^\ast}$, satisfies $(r, \xi)$-clustering. Then, any energy eigenstate, $\ket{E_m}$, satisfies
\begin{align}
  \hspace*{-0.3cm} 
   | \bra{E_m} (\Omega/n) \ket{E_m}\! -\! \langle \Omega /n \rangle_{\infty}  |  & \!  \le  \! ( 1/\sqrt{n} ) \max ( c_1 {\cal A}_1 , c_2  {\cal A}_2) ,  \notag   \\
    {\cal A}_1  =&  \left[ \beta^{\ast} E_m + \log Z_{\beta^{\ast}}  \right]^{(d+1)/2}  \!\! ,       \label{strong_ETH_low_energy}       \\    
    {\cal A}_2 =& \left[ \ell^d (\beta^{\ast}E_m + \log Z_{\beta^{\ast}}  ) \right]^{1/2}  \! , \nonumber
\end{align}
where $\langle \cdots \rangle_{\infty}$ is the average in the ground state and the constants $c_1,c_2$ depend on $d,r,\xi$, and $\ell_0$.
\end{theorem}
As we have set the energy of the ground state to zero, the quantity $\log Z_{\beta^{\ast}}$ always has a positive value~\cite{ft2}. We emphasize that even in the presence of degeneracy, the theorem is valid for an arbitrary superposition of degenerate eigenstates.

\begin{figure*}[t]
\centering
{\includegraphics[clip, scale=0.4]{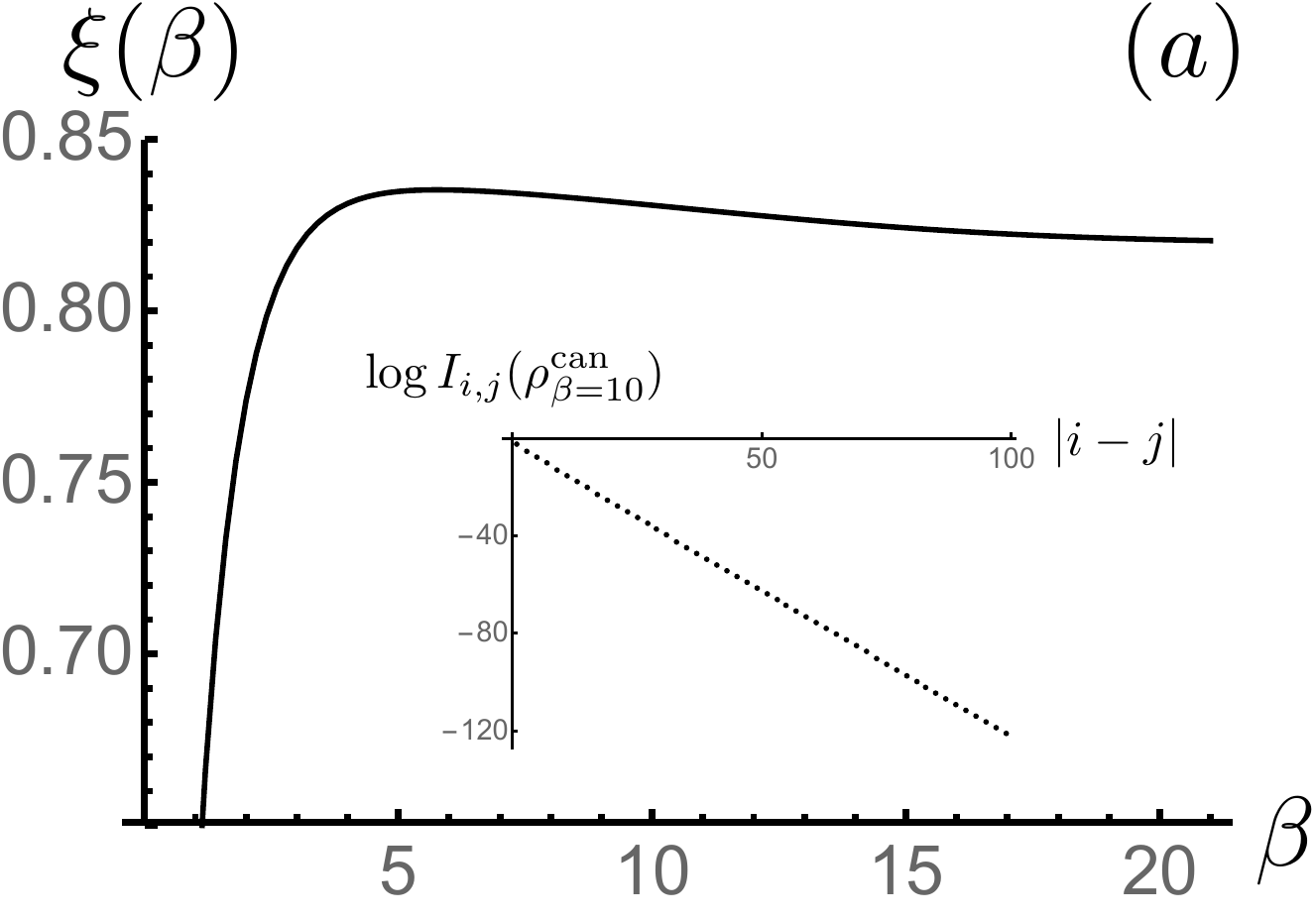}
}
{\includegraphics[clip, scale=0.4]{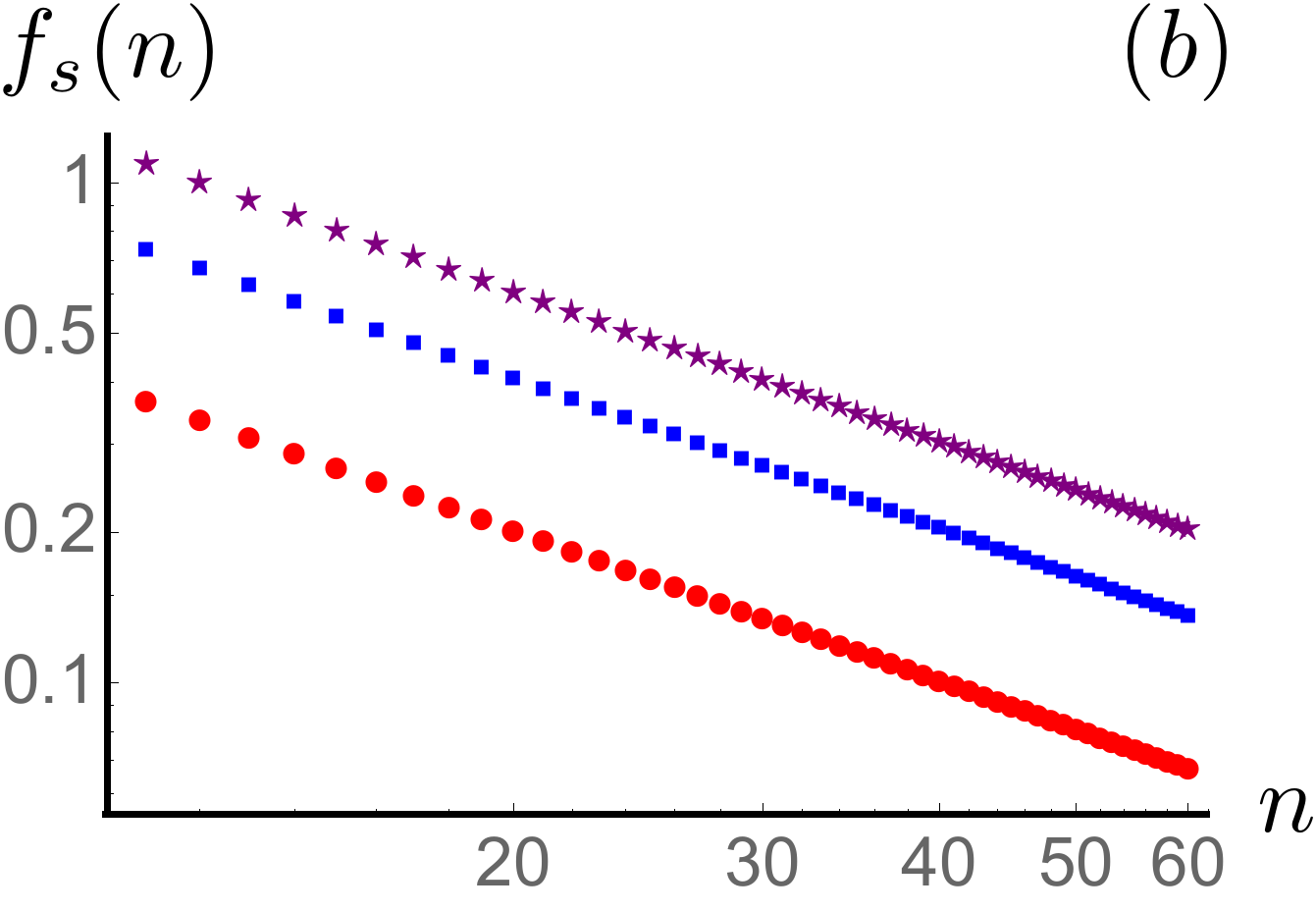}
}
{\includegraphics[clip, scale=0.4]{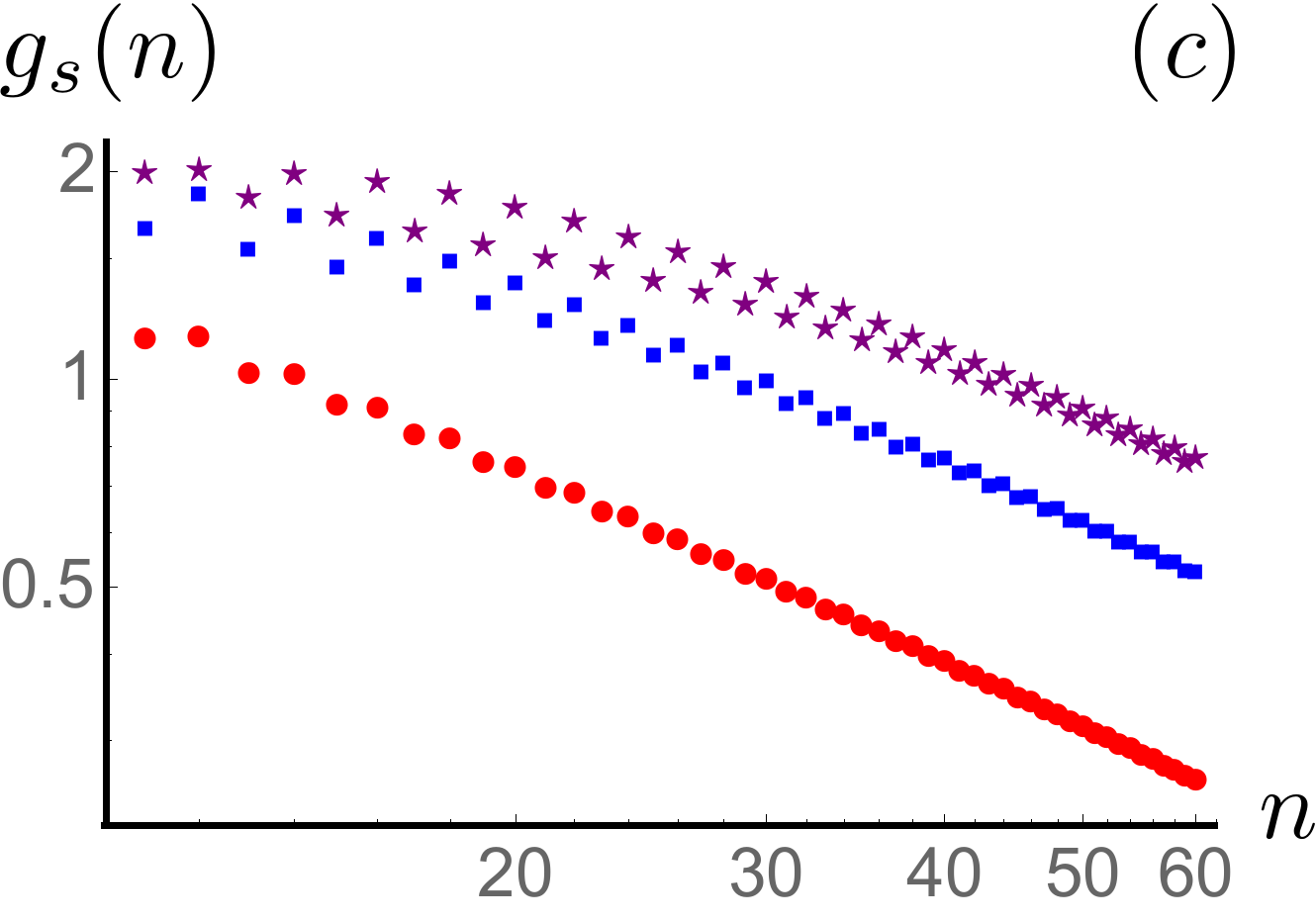}
}
\caption{(color online) Numerical demonstrations. The first figure (a) shows $\beta$-dependence of the correlation length, $\xi(\beta)$, in the canonical state $\rho_{\beta}$
($h=3/2$). 
We calculate the mutual information for the spin pairs of $\{r, n-r+1\}_{r=1}^{n/2}$ with $n=100$. In order to determine the correlation length, $\xi(\beta)$, we fit $\log [I_{i,j}(\rho_\beta)]$ with respect to $|i-j|$ by a linear function using the least squares method. The second and third figures (b) and (c) show the log--log plots of the system-size dependence of the functions, $f_s(n)$ and $g_s(n)$.
In the plots, points represented by circle ($\bullet$), square ({\tiny $\blacksquare$}), and star ($\star$) correspond to the cases of $s=3$, $s=5$, and $s=7$, respectively. For $s=7$, the behavior of $f_s(n)$ and $g_s(n)$ for a large $n$ are estimated by $n^{-0.998}$ and $n^{-0.854}$, respectively.
}
\label{fig:numerical_test}
\end{figure*}

If we consider the system that shows $\log Z_{\beta^{\ast}} \sim \orderof{1}$~\cite{ft3} and, in addition, look at the energy regime $E_m \lesssim \mathcal{O}(n^{1/(d+1)})$, the right-hand side vanishes in the thermodynamic limit. This implies the eigenstate thermalization. In other words, these low-lying eigenstates are indistinguishable from the ground state as long as the local observables are measured. A possible realistic situation for obtaining this scenario is given by the systems satisfying the following relation for the density of states in an energy shell $({{E}}-1,{{E}}]$ for the low-energy regime (See Fig.\ref{result_outline})~\cite{PhysRevB.76.035114, PhysRevB.92.115134,PhysRevA.80.052104}:
\begin{align}
{\cal N}_{{{E}}, 1} & \le n^{c {{E}} } \label{density_of_state_DE}  . 
\end{align}
In this case, we can demonstrate that the quantity $\log Z_{\beta^{\ast}}$ with $\beta^{\ast} \propto \log n$ becomes the order of $1$ in the thermodynamic limit~\cite{ft4}. 
Behavior of the density of states~\eqref{density_of_state_DE} is ubiquitous especially in the gapped systems as discussed in Refs.~\cite{PhysRevB.76.035114, PhysRevB.92.115134,PhysRevA.80.052104}. This behavior is rather general, when low-lying eigenstates are constructed with a superposition of locally excited states. We also demonstrate this with a specific model below. Notably, in the finite-dimensional gapped systems, it has been rigorously proved that the ground state shows an exponential decay for
spatially separated observables~\cite{ref:Hastings2006-ExpDec,Nachtergaele2006}, and hence, it is generally expected that the canonical distribution exhibits the clustering property in an extremely low-energy regime.

As the theorem \ref{main_theorem_Ensemble Equivalence_Strong} is derived from the clustering property, the scenario described here should be satisfied irrespective of the (non)integrability. This is quite different from the eigenstate thermalization observed in the high-energy regime, which generally requires the nonintegrability of the Hamiltonian.

\sectionprl{Numerical demonstration on the eigenstate thermalization }
To obtain a better understanding of the realization of eigenstate thermalization, we consider a simple integrable model. Let us consider the $xy$ model with a magnetic field:
\begin{align}
  H=\sum_{i=1}^{n-1}[ (3/4) \sigma_i^x   \sigma_{i+1}^x + (1/4) \sigma_i^y   \sigma_{i+1}^y] +   h \sum_{i=1}^n \sigma_i^z   , \label{XY_model_H}
\end{align}
where $\sigma_i^{\alpha} \, (\alpha=x,y,z)$ is the $\alpha$-component of the Pauli matrix at the site $i$. We consider the gapped case with $h>1$. 
The Hamiltonian is diagonalized into the Fermionic representation~\cite{PhysRevB.64.064412,LIEB1961407}, $H=\sum_{k=1}^n  2\epsilon_k c_k^{\dagger} c_k $, where $\epsilon_k$ is a positive eigenmode energy for the $k$th mode and $c_k^\dagger$ is the creation operator (we adjust the energy of the ground state to zero). An arbitrary eigenstate is expressed as
\begin{align}
  \ket{E_m} & = \prod_{k=1}^n[ c_k^\dagger]^{q_k} \ket{0} \quad q_k=0 \quad {\rm or} \quad 1  ,\label{eigen_states_XY} 
\end{align}
where $\ket{0}$ is the vacuum state and $m$ is determined by the choice of $\{q_k\}_{k=1}^n$. In this representation, the ground state is identical to the vacuum state. For $h>1$, each eigenmode energy, $\epsilon_k$, is of the order $1$. Hence, the number of eigenstates can be estimated through a simple combinatorial argument~\cite{ftc}, which justifies the relation~\eqref{density_of_state_DE}.

We here verify the clustering property in the present model. We consider the mutual information between two separated spins for $h=3/2$. 
For a given density matrix $\rho$, the mutual information $I_{i,j}(\rho)$ between the sites $i$ and $j$ is defined as, $I_{i,j}(\rho)=S(\rho_i)+S(\rho_j)-S(\rho_{i,j})$, where $S (\rho')$ is the Von Neumann entropy of the density matrix $\rho'$; and $\rho_{i}$ and $\rho_{i,j}$ are reduced density matrices of a single site, $i$, and two sites $i,j$, respectively. We first verify that $I_{i,j}(\rho_{\beta})$ shows an exponential decay as a function of distance $|i-j|$, which is demonstrated in the inset of Fig.~\ref{fig:numerical_test} (a). We calculate the $\beta$-dependence of the correlation length $\xi(\beta)$ and present a plot in the primary graph of Fig.~\ref{fig:numerical_test} (a). This plot indicates the clustering property even for extremely low-temperatures. 

From the clustering property and the energy dependence of the number of states \eqref{density_of_state_DE}, the emergence of the eigenstate thermalization can be directly observed. Let $\mathcal{E}_{s}$ be the set of all the eigenstates~\eqref{eigen_states_XY} with $\sum_{k=1}^n q_k\le s$. The set includes low-lying excited states whose energy is of the order $1$. Hence, we consider a deviation from the local thermodynamic property for the eigenstates in the $\mathcal{E}_{s}$ set and the ground state. We define the follwoing two quantities:  
\begin{align}
    f_{s}(n)    & : = \max_{\ket{E_m} \in \mathcal{E}_{s}} (1/n) | \bra{E_m} \hat{M}_z \ket{E_m} - \bra{E_0} \hat{M}_z \ket{E_0}| \, , \notag  \\
    g_{s}(n)    & : = \max_{\ket{E_m}  \in \mathcal{E}_{s}}  \bigl\|\rho^{[m]}_{i_0,i_0+1}- \rho^{[0]}_{i_0,i_0+1} \bigr\|_1 \, ,
             \label{func:Mz_local_norm}
\end{align}
where we use magnetization as a local observable, i.e., $\hat{M}_z=\sum_{i=1}^n \sigma_i^z$, and $i_0=\lfloor n/2\rfloor$. The matrices $\rho^{[m]}_{i_0,i_0+1}$ and $\rho^{[0]}_{i_0,i_0+1}$ are the reduced density matrices at the sites $\{i_0,i_0+1\}$ of the density matrix $\ket{E_m}\bra{E_m}$ and the ground state $\ket {E_0} \bra{E_0}$, respectively. 
The quantities $f_{s}(n)$ and $g_{s}(n)$ measure the deviation in terms of local observable (magnetization) and  the reduced density matrices, respectively, between eigenstates in the set $\mathcal{E}_{s}$ and the ground state.
In Fig.~\ref{fig:numerical_test} (b) and (c), we show $f_{s}(n)$ and $g_{s}(n)$ for $s=3,5,7$ as a function of the system size, $n$. 
The figures show the systematic decay of these quantities as the system size increases. This is a direct and clear indication of the emergence of eigenstate thermalization. 
The decay rate in this specific case is indeed faster by a factor of $n^{-1/2}$ compared to the decay rate obtained from the theorem \ref{main_theorem_Ensemble Equivalence_Strong}. 

\sectionprl{Results in the high-energy regime}
As the clustering property is satisfied even in integrable systems, the eigenstate thermalization defined in Eq.~\eqref{ee} should not be derived for the high energy regime, where the density of states is exponentially large with respect to the system size as depicted in Fig.~\ref{result_outline}. Instead, we prove the weak version of eigenstate thermalization.

We first derive a theorem on the ensemble equivalence between microcanonical and canonical ensembles, which shows that the ensemble equivalence holds with a small energy width, while it also shows that the eigenstate thermalization cannot be achieved in the high energy regime only from the clustering property. The theorem is given as follows. 
\begin{theorem} \label{thm:main_theorem_Ensemble Equivalence}
Let us consider the energy shell, $({E}-\Delta,{E}]$ and the inverse temperature, $\beta$ that maximizes $e^{-\beta {E}}{\cal N}_{{E},\Delta}$~\cite{Tasaki2018}.
If the canonical ensemble, $\rho_\beta$, satisfies the $(r, \xi)$-clustering, the microcanonical ensemble, $\rho_{{E},\Delta}$, satisfies the following: 
  \begin{align}
    \hspace*{-0.3cm}    
    \left|\langle \Omega /n \rangle_{{E},\Delta}   - \langle \Omega /n \rangle_{\beta} \right|  &\le  (1/\sqrt{n}) \max ( c'_1 {\cal B}_1 , c'_2 {\cal B}_2 ) , \notag  \\  
   {\cal B}_1  & =  \left[ \log (\sqrt{n} / \Delta ) \right]^{(d+1)/2}  ,  \label{thm1:op_ensemble_equiv}   \\    
   {\cal B}_2 & = \left[\ell^d  \log (\sqrt{n}/\Delta) \right]^{1/2}  , \nonumber
\end{align}
where $\langle \Omega  \rangle_{\beta }$ is the average with respect to the corresponding canonical ensembles, and the constants $c'_1$, $c'_2$ depend on $d$, $r$, $\xi$, and $\ell_0$.  
\end{theorem}
Based on the theorem, we can estimate a minimum value to justify the ensemble equivalence for the large system size. To obtain $\langle \Omega/n \rangle_{{E},\Delta} \approx  \langle \Omega/n \rangle_{\beta}$ for $n\to \infty$, we need to impose ${\cal B}_{1,2} \lesssim \orderof{n^{1/2}}$ from \eqref{thm1:op_ensemble_equiv}.
This condition gives the lower bound on the energy width, $ \Delta_c$, for the ensemble equivalence as \cite{ft1}
\begin{align}
  \Delta_c  &\sim \exp \left[-\orderof{n^{1/(d+1)}}\right]  . \label{delc}
\end{align}
This implies that even a subexponentially small energy width is sufficient to guarantee the ensemble equivalence.
Although studies on the ensemble equivalence has a long history~\cite{Lima1972,lima1971equivalence,de2006quantum,Muller2015,Georgii1995,Touchette_2011,Touchette2015}, the finite-size effect has been studied quite recently \cite{brandao2015equivalence,Tasaki2018}.
Our system-size dependence is an exponential improvement from the state-of-the-art estimation $\orderof{n^{-1/2}}$ in~\cite{Tasaki2018}.
However, even this energy width cannot discriminate a single eigenstate in the high energy regime having exponentially high density of states with respect to the system size, and hence one cannot reach the eigenstate thermalization.

We next show that the weak version of the eigenstate thermalization can be quantitatively argued~\cite{PhysRevLett.105.250401,PhysRevLett.119.100601}. As defined in \eqref{wee}, the weak version of eigenstate thermalization states that almost all the eigenstates in the energy shell have the same property. Using the variance, ${\rm Var}_{{E},\Delta}(\Omega/n)$, defined in Eq.~\eqref{wee_var}, we can derive the following finite-size effect (see Sec.~III in \cite{Supplement_ETH_clustering}):
\begin{align}
  {\rm Var}_{{E},\Delta} (\Omega/n)  &\le  (1/n) \max ( c''_1 {\cal B}_1^2 , c''_2 {\cal B}_2^2 )  \, , 
  \label{weak_eth_finite0}
\end{align}
where $ {\cal B}_1,  {\cal B}_2$ were defined in Eq.~\eqref{thm1:op_ensemble_equiv}, and $c_1''$, $c_2''$ are constants that depend on $d$, $r$, $\xi$, and $\ell_0$. The inequality~\eqref{weak_eth_finite0} implies that by taking the energy width $\Delta=1/\poly(n)$, the following holds 
\begin{align}
  {\rm Var}_{{E},\Delta} (\Omega/n)  & \lesssim \log^{d+1} (n) /n \, . \label{weak_eth_finite}
\end{align}                         
It is noteworthy that recent calculations by Alba~\cite{PhysRevB.91.155123} showed an example that expresses the variance of $\orderof{1/n}$. This indicates that our estimation~\eqref{weak_eth_finite} is the best general upper bound on the weak ETH up to a logarithmic correction.

\sectionprl{Summary and discussion}
Theorem~\ref{main_theorem_Ensemble Equivalence_Strong} gives a clear scenario for achieving the eigenstate thermalization in low-lying eigenstates near the ground state, irrespective of the (non)integrability of the Hamiltonian. This is quite different from the eigenstate thermalization in the high-energy regime where the nonintegrability is assumed to be a key-ingredient~\cite{PhysRevLett.98.050405,PhysRevLett.106.227203}.
We also emphasize that the ETH can be argued at the level of reduced density matrix for local sites, which is presented in Sec. VI in~\cite{Supplement_ETH_clustering}. 
Moreover, our approach on ensemble equivalence can be extended to a wider class of systems for e.g., long-range interacting systems where the Chernoff-Hoeffding type concentration bound still holds~\cite{Kuwahara_2016_gs, kuwa2017,kuwahara2019gaussian}

We comment on the relation between the present argument and the many-body localization phenomenon~\cite{doi:10.1146/annurev-conmatphys-031214-014726}, where the ETH is violated. In the present analysis, we consider sufficiently low-energy regime, $\beta^{\ast} \propto \log n$. Although the one-dimensional systems exhibit the clustering property in general~\cite{Araki1969}, it is justified only for $\beta \lesssim\orderof{1}$. In systems with many-body localization, the canonical state $\rho_{\beta}$ does not satisfy the clustering property for $\beta^{\ast} \propto \log n$. We have discussed this point in Sec. VI of~\cite{Supplement_ETH_clustering} with a numerical calculation. We stress that Theorem~\ref{main_theorem_Ensemble Equivalence_Strong} has no inconsistency with the absence of thermalization in systems with many-body localization.

\begin{acknowledgments}
The work of T. K. was supported by the RIKEN Center for AIP and JSPS KAKENHI Grant No. 18K13475.
TK gives thanks to God for his wisdom.
K.S. was supported by JSPS Grants-in-Aid for Scientific Research (JP16H02211, JP19H05603).
\end{acknowledgments}

\bibliography{ETH}

\begin{thebibliography}{69}
\bibitem [{ft2()}]{ft2}%
  \BibitemOpen
  \href@noop {} {}If we consider a general value for the energy of the ground
  state, the same relation can be derived by shifting the eigenenergy $E_m$ to
  $E_m-E_0$\BibitemShut {NoStop}%
\bibitem [{ft3()}]{ft3}%
  \BibitemOpen
  \href@noop {} {}More precisely, $\log Z_{\beta^{\ast}} \sim n^{\alpha}$
  $(\alpha <1/2)$ is sufficient for convergence in the right-hand
  side\BibitemShut {NoStop}%
\bibitem [{ft4()}]{ft4}%
  \BibitemOpen
  \href@noop {} {}For ${\cal N}_{E, 1} \le n^{c E}$ we assume $\beta^{\ast} =
  c^{\ast} \log n$ satisfying $c^{\ast} > c$. Subsequently, we have
  $Z_{\beta^{\ast}} =\sum_{m=0}^{2^{n}-1} e^{-\beta^{\ast} E_m}\le 1+\sum_{j\ge
  1} (n^{c-c^{\ast}})^j$, which is finite even for the thermodynamic limit.
  Recall that the energy of the ground state is set to zero. Note that in the
  high-energy regime, the bound \eqref{density_of_state_DE} is too loose to
  describe the true system-size dependence of ${\cal N}_{E,1}$, and hence this
  computation overestimates the amplitude of $Z$ (nevertheless, it is still of
  $\orderof{1}$)\BibitemShut {NoStop}%
\bibitem [{ftc()}]{ftc}%
  \BibitemOpen
  \href@noop {} {}Note that $2 \epsilon_k $ for all $k$ are larger than the
  energy gap. Then, let us define $\tilde{\epsilon}$ as such $2\epsilon_k \ge
  \tilde{\epsilon}$ for all $1\le k\le n$. Subsequently, the total number of
  eigenstates whose energy is smaller than $\bar{q}\tilde{\epsilon}$ is less
  than $\sum_{q\le \bar{q}} \binom{n}{q} \le n^{\bar{q}}$. This implies ${\cal
  N}_{E, E} \le n^{E/\tilde{\epsilon}}$, and hence ${\cal N}_{E, 1} \le
  n^{E/\tilde{\epsilon}}$\BibitemShut {NoStop}%
\bibitem [{ft1()}]{ft1}%
  \BibitemOpen
  \href@noop {} {}From the inequalities on ${\cal A}_1$ and ${\cal A}_2$, we
  can derive conditions $\Delta \ge \sqrt{n} e^{-n^{1/(d+1)}}$ and $\Delta \ge
  \sqrt{n} e^{-n/\ell^d}$, respectively. Since for large $n$ the former
  inequality is stronger than the latter one, a general lower bound of the
  energy width is given by \eqref{delc}\BibitemShut {NoStop}%
\end{thebibliography}

\clearpage
\newpage

\begin{widetext}
\begin{center}
{\large \bf Supplementary Material for  \protect \\ 
  ``Ensemble equivalence and eigenstate thermalization from clustering of correlation'' }\\
\vspace*{0.3cm}
Tomotaka Kuwahara$^{1,2}$ and Keiji Saito$^{3}$ \\
\vspace*{0.1cm}
$^{1}${\small \it Mathematical Science Team, RIKEN Center for Advanced Intelligence Project (AIP),1-4-1 Nihonbashi, Chuo-ku, Tokyo 103-0027, Japan \protect \\
$^{2}$Interdisciplinary Theoretical \& Mathematical Sciences Program (iTHEMS) RIKEN 2-1, Hirosawa, Wako, Saitama 351-0198, Japan} \\
$^{3}${\small \it Department of Physics, Keio University, Yokohama 223-8522, Japan} 
\end{center}

\setcounter{equation}{0}
\renewcommand{\theequation}{S.\arabic{equation}}


\section{Setup}
\subsection{Notations and definitions}
\noindent
We iterate several definitions that are presented in the primary text.
The Hamiltonian that we considered consists of $n$ local terms: 
\begin{align}
H= \sum_{i=1}^n H_i  \, ,\quad \|H_i\| \le 1. 
\label{Hamiltonian_supple}
\end{align}
The local Hamiltonian $H_i$ contains spin operators that act nontrivially on spins $j$ with the distance $\dist(i,j)\le \ell_0$.

\vspace*{0.5cm}
\noindent
The canonical and microcanonical distributions are defined as
\begin{align}
\rho_{\beta} & := {e^{-\beta H} \over  Z_{\beta}} \, ,  \\
\rho_{{{E}},\Delta}& :={ 1 \over  {\cal N}_{{{E}},\Delta}} \sum_{E_m \in ({{E}}- \Delta, {{E}} ] } \ket{E_m}\bra{E_m} \, ,
\end{align}
where $Z_{\beta}$ is the partition function $Z_{\beta} =\tr e^{-\beta H}$, and $\ket{E_m}$ is the $m$th eigenstate of the Hamiltonian ($m=0,1,\ldots,2^n-1$) and we define
\begin{align}
{\cal N}_{{{E}},\Delta} & := \tr \left[ \sum_{E_m \in ({{E}}- \Delta, {{E}} ] } \ket{E_m}\bra{E_m}\right]  \, .
\end{align}
For an arbitrary operator $O$, the canonical and microcanonical averages are expressed as follows, respectively:
\begin{align}
  \langle O \rangle_{\beta} &  := \tr (  \rho_{\beta} O) \, , \\
  \langle O\rangle_{{{E}},\Delta} & :=\tr( \rho_{{{E}},\Delta} O) \, .
\end{align}

\vspace*{0.5cm}
\noindent
Next, we focus on the following operator:
 \begin{align}
   \Omega   = \sum_{i=1}^n \Omega_i  ,\quad \|\Omega_i\| \le 1,\label{Def:Omega_l_local}
\end{align}
where $\Omega_i$ is composed of spin operators that act on spins $j$ with the distance $\dist(i,j)\le \ell$. 

\vspace*{1cm}
\noindent
In our theory, we assume the clustering property for $\rho_{\beta}$, which is defined as follows:
  \begin{definition} \label{Def:main_theorem_Ensemble Equivalence_re}
    {\it Let $A_X,B_Y$ be arbitrary operators supported on subsets $X$ and $Y$, respectively. 
We say that a density matrix $\rho$ satisfies the $(r, \xi)$-clustering if 
 \begin{align}
\left| \tr(\rho A_X B_Y) - \tr(\rho A_X )\tr(\rho B_Y)\right| \le \|A_X\| \|B_Y\| e^{-\dist(X,Y)/\xi} \notag 
\end{align}
for $\dist(X,Y)\ge r$, where $r$ and $\xi$ are fixed constants.}
\end{definition}

\subsection{On the definition of eigenstate thermalization and ensemble equivalence}
We start with the established definition on ensemble equivalence. Suppose that we have two ensembles $\rho_A$ and $\rho_B$. We say that the ensemble equivalence holds between these ensembles in terms of the observable $O$, if the following is satisfied
\begin{align}
  {\rm Tr}(\rho_A O ) - {\rm Tr} (\rho_B O) &= 0 \, .
\end{align}
Examples include the ensemble equivalence for the canonical ensemble $\rho_{\beta}$ and the microcanonical ensemble $\rho_{E,\Delta}$ as we address in this paper. Another example is the equivalence between the canonical ensemble for a many-particles system and the Gibbs ensemble. In this paper, we are interested in the local observable $\Omega /n$. Then, the ensemble equivalence between $\rho_{\beta}$ and $\rho_{E,\Delta}$ holds if the following is satisfied
\begin{align}
  {\rm Tr}(\rho_{E,\Delta} (\Omega / n ) ) - {\rm Tr} (\rho_{\beta} (\Omega / n)) &= 0 \, .
\end{align}
As an extreme case, we consider very small energy window so that the energy shell contains a single eigenstate $|E_{m}\rangle$. In this case, the microcanonical ensemble is replaced by $| E_{m} \rangle \langle E_{m}|$ leading to
\begin{align}
  {\rm Tr}(|E_{m}\rangle \langle E_{m} | (\Omega / n ) ) - {\rm Tr} (\rho_{\beta} (\Omega / n)) &= 0 \, .
\end{align}
This is the eigenstate thermalization meaning that the single eigenstate is equivalent to the canonical ensemble in terms of the local observable.

\subsection{Concentration inequality}
\noindent
As the core of the proof of our theorems, we utilize the following concentration inequality that is resulted from the $(r, \xi)$-clustering.
We start with the following lemma, whose proof is shown in Section~\ref{Sec:Concentration bound from the clustering}.
This lemma is essentially the same as that in Ref.~\cite{Anshu_2016}.
\begin{lemma}  \label{lem:main_lemma_Anshu}
  {\it 
Let $\rho$ be an arbitrary density matrix that satisfies the $(r, \xi)$-clustering (Def.~\ref{Def:main_theorem_Ensemble Equivalence_re}).
Subsequently, for an arbitrary even integer $M$, the $M$-th moment of the operator $\Omega$ is bounded from above by
 \begin{align}
\tr \Bigl [ \bigl(   \Omega-\langle    \Omega \rangle_\rho\bigr)^M \rho \Bigr] 
&\le [8\alpha M n   (3\ell)^d]^{M/2}  +  3\xi [8\alpha  n M^{d+1} (3\xi d /2)^d]^{M/2}  \notag \\
&=:  3\xi [\tilde{c}_1 n M^{d+1}/(2e)]^{M/2} + [\tilde{c}_2\ell^d M n/(2e)]^{M/2}  \label{moment_bound_concentration}
\end{align}
with $\langle \cdots \rangle_\rho:= \tr(\rho \cdots)$ and 
 \begin{align}
\tilde{c}_1:=16e \alpha  (3\xi d /2)^d   \quad {\rm and} \quad \tilde{c}_2:= 3^d(16e\alpha) , 
\end{align}
where $\alpha$ is a geometric parameter defined in Ineq.~\eqref{def:X_i^s_alpha} of Section~\ref{Sec:Concentration bound from the clustering}, and we assume $\ell \ge r$.  If $\ell \le r$, the parameter $\ell$ in \eqref{moment_bound_concentration} is replaced by $r$. }
\end{lemma}

\vspace*{1cm}
\noindent
We define the probability density to observe the value $x$ for the observable $\Omega$ in terms of the density matrix $\rho$
\begin{align}
P_{\rho} (x)  &:= \tr \left[ \rho \delta ( x - \Omega ) \right] \, , \label{pdensity}
\end{align}
where $\delta (\cdots )$ is the Dirac delta function. We relate the inequality~\eqref{moment_bound_concentration} to the concentration bound for this distribution.
For the given positive value $x_0$, the following relation for the cumulative distribution is satisfied:
\begin{align}
  P_{\rho}(|x-\langle \Omega \rangle_\rho| \ge x_0 ) := \int_{|x-\langle \Omega \rangle_\rho| \ge x_0 } P_{\rho}(x) d x  
  &= P_{\rho}((x-\langle \Omega \rangle_\rho)^M \ge x_0^M )    \nonumber \\
  &\le  { \tr \left[ \rho (x-\langle \Omega \rangle_\rho)^M \right] \over  x_0^M }
    \nonumber \\
  & \le    3\xi \left(\frac{\tilde{c}_1 n M^{d+1}}{2e x_0^2} \right)^{M/2} +  \left(\frac{\tilde{c}_2\ell^d M n}{2ex_0^2} \right)^{M/2} \, ,\label{bound_concent_1}
\end{align}
where $M$ is a positive even integer. We use the Markov inequality for the second line, and the inequality (\ref{moment_bound_concentration}) for the third line. We subsequently choose $M/2$ ($\in \mathbb{Z}$) as 
 \begin{align}
\frac{M}{2}=\left \lfloor \min \left\{ \left(\frac{x_0^2}{\tilde{c}_1 n} \right)^{1/(d+1)} ,~\frac{x_0^2}{\tilde{c}_2\ell^d n} \right\} \right\rfloor , \label{choice_of_m_para}
 \end{align}
 where $\lfloor \cdots \rfloor$ is the floor function. 
This choice ensures $\frac{\tilde{c}_1 n M^{d+1}}{2ex_0^2}\le 1/e$ and $\frac{\tilde{c}_2\ell^d M n}{2e x_0^2} \le 1/e$; hence, the upper bound \eqref{bound_concent_1} reduces to
\begin{align}
  P_{\rho}(|x-\langle \Omega \rangle_\rho| \ge x_0 ) &\le  (e+3e\xi)  \max \left( e^{-[x_0^2/(\tilde{c}_1 n)]^{1/(d+1)}}  ,  e^{-x_0^2/(\tilde{c}_2\ell^d n)} \right) \, .
\end{align}
Note that the cumulative distribution must be less than $1$. Combining this with the inequality above, we obtain the following relation:
\begin{align}
  P_{\rho}(|x-\langle \Omega \rangle_\rho| \ge x_0 ) &\le
\min \left[ 1, (e+3e\xi)  \max \left( e^{-[x_0^2/(\tilde{c}_1 n)]^{1/(d+1)}}  ,  e^{-x_0^2/(\tilde{c}_2\ell^d n)} \right) \right]. \label{concentration_bound_clustering}
\end{align}

\subsection{Useful formula}
\noindent
For the calculations in the following sections, the following lemma is useful.
\begin{lemma} \label{Prob:expectation}
{\it Let $p(x)$ be an arbitrary probability distribution whose cumulative distribution is bounded from above:
\begin{align}
  P(| x-a |\ge x_0):= \int_{|x-a| \ge x_0}p(x) dx \le \min(1, e^{-x_0^\gamma / \sigma + x_1 }) \, , ~~~~~\gamma >0, ~\sigma >0 ,~x_0>0\, .
\end{align}
Subsequently, for an arbitrary $k \in \mathbb{N}$, we obtain 
\begin{align}
\int_{-\infty}^\infty |x-a|^k  p(x) dx \le  (2 \sigma x_1)^{k/\gamma} + \frac{k}{\gamma} (2\sigma)^{k/\gamma} \Gamma(k/\gamma)   \label{ineq:Prob:expectation}
\end{align}
with $\Gamma (\cdot)$ as the gamma function.}
\end{lemma}

\subsubsection{Proof of Lemma~\ref{Prob:expectation}}
\noindent
We first define $\tilde{x}:= (2\sigma x_1)^{1/\gamma}$, which is the solution of $-\tilde{x}^\gamma / \sigma+ x_1 =  -\tilde{x}^\gamma /(2 \sigma )$.
Subsequently, $-x_0^\gamma / \sigma + x_1 \le -x_0^\gamma /(2 \sigma)$ for $x_0 \ge \tilde{x}$; that is, $P(|x-a|\ge x_0) \le e^{-x_0^\gamma /(2 \sigma)}$ for  $x_0 \ge \tilde{x}$. Using this, we obtain the following relation:
\begin{align}
\int_{-\infty}^\infty |x-a|^k p(x) dx =\int_{-\infty}^\infty |x|^k p(x+a) dx
&\le  \tilde{x}^k\int_{-\tilde{x}}^{\tilde{x}} p(x+a) dx  + \int_{x \ge \tilde{x}} x^k [p(x+a)+p(-x+a)] dx \nonumber \\
                                      &\le \tilde{x}^k - \int_{\tilde{x}}^\infty x^k \frac{d}{dx} \left[ \int_{|x' -a|\ge x} p(x') d x'  \right]  dx . \label{the_ineq_expec_dist1}
\end{align}
The second term reduces to
\begin{align}
  -\int_{\tilde{x}}^\infty x^k  \frac{d}{dx} \left[ \int_{|x' -a|\ge x} p(x') d x'  \right]  dx&  =-\tilde{x}^k P(|x-a| \ge \tilde{x}) + \int_{\tilde{x}}^\infty k x^{k-1}
\left[ \int_{|x' -a|\ge x} p(x') d x'  \right]    dx  \notag \\
&\le  \int_{0}^\infty k x^{k-1} e^{-x^\gamma /(2 \sigma)} dx  
=  \frac{k}{\gamma} (2\sigma)^{k/\gamma} \Gamma(k/\gamma)  , \label{the_ineq_expec_dist2}
\end{align}
where we use the partial integration in the first equation.
By combining the two inequalities~\eqref{the_ineq_expec_dist1} and \eqref{the_ineq_expec_dist2}, we obtain the inequality~\eqref{ineq:Prob:expectation}.
This completes the proof. $\square$

\section{Proof of Theorem~\ref{main_theorem_Ensemble Equivalence_Strong}   in the main text.}
\noindent
In this section, we prove Theorem~\ref{main_theorem_Ensemble Equivalence_Strong}  .  We start by defining the probability distribution to observe the value $x$ for the observable $\Omega$ for a given eigenstate $|E_m\rangle$:
\begin{align}
  P_{E_m} (x)  &= \tr \left[ \,  | E_m \rangle \langle E_m | \, \delta (x - \Omega )  \,  \right] \, .
\end{align}
Subsequently, we consider the cumulative distribution $ P_{E_m}(|x - \langle \Omega \rangle_{\beta^{\ast}}  | \ge x_0)$ where, at this stage, the inverse temperature $\beta^{\ast}$ is arbitrary. From Markov's inequality with a positive even integer $M$, we have the following relation:
\begin{align}
  P_{E_m}( |x - \langle \Omega \rangle_{\beta^{\ast}}| \ge x_0 ) &= P_{E_m}( (x - \langle \Omega \rangle_{\beta^{\ast}})^M \ge x_0^M  )     \nonumber \\
                                                               & \le   {\bra{E_m} (\Omega  -  \langle \Omega \rangle_{\beta^{\ast}}                                                                 )^M \ket{E_m} \over  x_0^M  }\,   . \label{Markov_Strong_ETH}
\end{align}
Below, we bound $\bra{E_m} (\Omega -  \langle \Omega \rangle_{\beta^{\ast}})^M \ket{E_m}$ from above using $\langle  (\Omega -  \langle \Omega \rangle_{\beta^{\ast}})^M \rangle_{\beta^\ast}$.
For notational convenience, we define $E_{\beta^\ast} := - (1/\beta^\ast)\ln Z_{\beta^\ast}$. Note that the operator $(\Omega  - \langle \Omega \rangle_{\beta^{\ast}}) ^M$ is a non-negative operator; hence, we can obtain the following relation:
\begin{align}
  \bra{E_m}  (\Omega -  \langle \Omega  \rangle_{\beta^{\ast}}) ^M
  \ket{E_m} &=e^{\beta^\ast E_m}  e^{-\beta^\ast E_m} \bra{E_m}  (\Omega  - \langle \Omega \rangle_{\beta^{\ast}}) ^M  \ket{E_m}\notag \\
          &\le  Z_{\beta^{\ast}} e^{\beta^\ast E_m}  \sum_{E_{m'} \in ( -\infty , \infty ) } { e^{-\beta^\ast E_{m'}} \over Z_{\beta^{\ast}}} \bra{E_{m'}} (\Omega -  \langle \Omega \rangle_{\beta^{\ast}}) ^M    \ket{E_{m'} } \notag \\
          & = e^{\beta^\ast (E_m-E_{\beta^\ast})} \langle (\Omega -     \langle \Omega \rangle_{\beta^{\ast}}) ^M \rangle_{\beta^\ast} \, .                                
\label{mc_Can_difference_Z2_ETH}
\end{align}
Combining the inequality~\eqref{Markov_Strong_ETH} with this, the following relation is obtained: 
\begin{align}
  P_{E_m}( |x -  \langle \Omega \rangle_{\beta^{\ast}}  | \ge x_0) & \le \,   e^{\beta^\ast (E_m-E_{\beta^\ast})}  {\langle (\Omega -  \langle \Omega \rangle_{\beta^{\ast}})^M \rangle_{\beta^\ast} \over x_0^M }\,
  . \label{P_tilde_Omega_bound2}
\end{align}
From the assumption of the ($r$,$\xi$)-clustering for the density matrix $\rho_{\beta^\ast}$, the $M$-th moment $\langle (x -  \langle \Omega \rangle_{\beta^{\ast}})^M  \rangle_{\beta^\ast}$ obeys the inequality~\eqref{moment_bound_concentration} from Lemma~\ref{lem:main_lemma_Anshu}.
Hence, we have
\begin{align}
  P_{E_m}(|x -  \langle \Omega \rangle_{\beta^{\ast}}  |\ge x_0)  &  \le   e^{\beta^\ast (E_m-E_{\beta^\ast})} \left[ 3\xi \left(\frac{\tilde{c}_1 n M^{d+1}}{2e x_0^2} \right)^{M/2} +  \left(\frac{\tilde{c}_2\ell^d M n}{2ex_0^2} \right)^{M/2} \right] \,    . \label{P_tilde_Omega_bound_2_1}
\end{align}
By choosing $M$ as in Eq.~\eqref{choice_of_m_para}, we arrive at the concentration bound as
\begin{align}
    P_{E_m}(|x -  \langle \Omega \rangle_{\beta^{\ast}}|\ge x_0) \le  \min\left\{ 1, \, e^{\beta^\ast (E_m-E_{\beta^\ast})} (e+3e\xi) \max\left( e^{ -[ x_0^2/(\tilde{c}_1 n)]^{1/(d+1)}}, e^{-x_0^2/(\tilde{c}_2n)}\right) \right\} . 
\label{P_omega_E_upper_bound}
\end{align}

\vspace*{0.5cm}
\noindent
Having obtained the concentration inequality, we evaluate the integration bound: 
\begin{align}
  |\bra{E_m}  \Omega \ket{E_m} -  \langle \Omega \rangle_{\beta^{\ast}}  | &= \left|  \int_{-\infty}^\infty ( x -  \langle \Omega \rangle_{\beta^{\ast}}  )    P_{E_m}(x)dx \right| \le   \int_{-\infty}^\infty | x -  \langle \Omega \rangle_{\beta^{\ast}}  |    P_{E_m}(x)dx \, .                                                                         \label{upper_bound_omega_l_E}
\end{align}
For evaluating the most right expression in (\ref{upper_bound_omega_l_E}), we use Lemma~\ref{Prob:expectation} with $k=1$. From the inequality given in \eqref{P_omega_E_upper_bound}, the parameter sets in Lemma~\ref{Prob:expectation} are identified as
\begin{align}
  a & =  \langle \Omega \rangle_{\beta^{\ast}}      \, ,  \nonumber \\
  \{\gamma, \sigma, x_1\} & = \{2/(d+1), (\tilde{c}_1n)^{\frac{1}{d+1}}, \beta^\ast (E_m-E_{\beta^\ast})+\log (e+3e\xi)  \} \, \quad {\rm and} \quad
   \{2, \tilde{c}_2\ell^d n, \beta^\ast (E_m-E_{\beta^\ast})+\log (e+3e\xi) \}. \notag 
\end{align}
We subsequently obtain the following from Ineq~\eqref{ineq:Prob:expectation}: 
\begin{align}
  |\langle \Omega \rangle_{E_m}  - \langle \Omega \rangle_{\beta^{\ast}}  |\le \max ( \tilde{\mathcal{A}}_1,\tilde{\mathcal{A}}_2  ) \, , 
\end{align}
where
\begin{align}
&\tilde{\mathcal{A}}_1:=\sqrt{\tilde{c}_1 n} \left[2\beta^\ast (E_m-E_{\beta^\ast})+2\log (e+3e\xi)\right]^{\frac{d+1}{2}}+2^{(d-1)/2} (d+1)\sqrt{\tilde{c}_1 n}\Gamma[(d+1)/2],   \notag \\
&\tilde{\mathcal{A}}_2:=\sqrt{2\tilde{c}_2\ell^d n \left[2\beta^\ast (E_m-E_{\beta^\ast})+2\log (e+3e\xi)\right]}+\frac{1}{2}\sqrt{2\pi \tilde{c}_2\ell^d n }   .\label{upp_bound_estimation_average_ensemble2}
\end{align}
Therefore, the upper bound \eqref{upper_bound_omega_l_E} reduces to 
\begin{align}
{1\over n} | \bra{E_m} \Omega \ket{E_m} - \langle \Omega  \rangle_{\beta^\ast}|
  \le {1 \over \sqrt{n}} \max\left( c_1[\beta^\ast (E_m- E_{\beta^\ast})]^{ (d+1)/2} , c_2[ \ell^d \beta^\ast (E_m- E_{\beta^\ast})]^{1/2}  \right),
\label{strong_ETH_low_energy_beta_ast}
\end{align}
where $c_1,c_2$ are constants that only depend on $d$, $\xi$, and $\alpha$.
Note that $|\bra{E_m} \Omega \ket{E_m}| = | \bra{E_m} \Omega \ket{E_m} - \langle \Omega  \rangle_{\beta^\ast}|$ for $ \langle \Omega  \rangle_{\beta^\ast}=0$.
This completes the proof of Theorem~\ref{main_theorem_Ensemble Equivalence_Strong}  . $\square$ 

\vspace*{1cm}
\noindent
We remark that the inequality~\eqref{strong_ETH_low_energy_beta_ast} also holds for the ground state:
\begin{align}
  \frac{1}{n}|  \bra{E_0}  \Omega  \ket{E_0}  - \langle \Omega  \rangle_{\beta^\ast}|
  \le {1\over \sqrt{n} } \max\left(c_1[\beta^\ast (- E_{\beta^\ast})]^{(d+1)/2} ,c_2[\ell^d \beta^\ast (- E_{\beta^\ast})]^{1/2}  \right) \, , 
\label{strong_ETH_low_energy_beta_ast_gs}
\end{align}
where $\ket{E_0}$ is the ground state. Recall that we set the ground energy equal to zero.
By combining the inequalities~\eqref{strong_ETH_low_energy_beta_ast} and \eqref{strong_ETH_low_energy_beta_ast_gs} with the triangle inequality, we have
\begin{align}
  \frac{1}{n}| \bra{E_m} \Omega \ket{E_m} -  \bra{E_0}  \Omega  \ket{E_0}
  | &\le
 \frac{1}{n}| \bra{E_m} \Omega \ket{E_m} - \langle \Omega \rangle_{\beta^\ast}|  +  \frac{1}{n}|   \bra{E_0}  \Omega  \ket{E_0} - \langle \Omega \rangle_{\beta^\ast}  | \notag \\
 &\le {2 \over \sqrt{n}} \max\left( c_1[\beta^\ast (E_m- E_{\beta^\ast})]^{ (d+1)/2} , c_2[ \ell^d \beta^\ast (E_m- E_{\beta^\ast})]^{1/2}  \right). \label{ineq_for_g_s_ETH}
\end{align}
When the ground states exhibit a $D$-fold degeneracy, we denote the ground states by $\{\ket{E_m}\}_{m=0}^{D-1}$ (i.e., $H\ket{E_m}=0$ for $j\le D-1$) and consider
\begin{align}
\langle \Omega \rangle_{\infty}=\frac{1}{D} \sum_{m=0}^{D-1} \bra{E_m}  \Omega \ket{E_m}  .
\end{align}
We subsequently obtain
\begin{align}
|\langle \Omega \rangle_{\infty} - \langle \Omega \rangle_{\beta^{\ast}}| &=\frac{1}{D}  \left|  \sum_{m=0}^{D-1} \bigl(\bra{E_m}  \Omega \ket{E_m} - \langle \Omega \rangle_{\beta^{\ast}}\bigr) \right|  \le \frac{1}{D}    \sum_{m=0}^{D-1}  \left| \bra{E_m}  \Omega \ket{E_m} - \langle \Omega \rangle_{\beta^{\ast}} \right| .
\end{align}
Therefore, we can derive the same inequality as \eqref{ineq_for_g_s_ETH} for the degenerate ground states. 


\section{Proof of Theorem~\ref{thm:main_theorem_Ensemble Equivalence}   in the main text.}
\noindent
The proof is almost similar to that of Theorem~\ref{main_theorem_Ensemble Equivalence_Strong}  .
In this section, we first prove the following inequality~\eqref{CH_micro_canonical} which implies that almost all the eigenstates in the energy shell $({{E}}- \Delta, {{E}}]$ have the same expectation value of $\langle \Omega \rangle_{\beta}$.
Then, the inequality~(13) in the main text is derived as the corollary.

Here, following Ref.~\cite{Tasaki2018}, we relate the microcanonical energy ${{E}}$ to the inverse temperature $\beta$ as follows:
\begin{align}
{{E}} = \arg\max_{E \in \mathbb{R}}\left( e^{-\beta E}{\cal N}_{E ,\Delta}\right) . \label{Def:U_micro_en_re}
\end{align}
We define a probability distribution $P_{{{E}},\Delta}(x)$ such that
\begin{align}
P_{{{E}},\Delta}(x) = {1\over  {\cal N}_{{{E}},\Delta}} \sum_{E_m \in ({{E}}- \Delta, {{E}} ] }  \delta (x -  \bra{E_m} \Omega\ket{E_m}) \, , \label{def:P_omega_x}
\end{align}where $\delta $ is the Dirac delta function. Using Lemma~\ref{lem:main_lemma_Anshu}, we obtain the following proposition (the proof is presented in the two subsequent subsections):
\begin{prop} \label{main_theorem_Ensemble Equivalence_clustering}
{\it  Let $P_{{{E}},\Delta}(|x - \langle \Omega \rangle_{\beta} | \ge x_0 )  \, (x_0>0)$ be the cumulative probability distribution of the distribution $P_{{{E}},\Delta}(x) $. Subsequently, if the canonical distribution with the corresponding inverse temperature $\rho_\beta$ exhibits the $(r, \xi)$-clustering property, we have
\begin{align}
  P_{{{E}},\Delta}\left(|x - \langle\Omega \rangle_{\beta}| \ge  x_0  \right) \le
\min\left[ 1, C_\Delta \max\left( e^{ -[ x_0^2/(\tilde{c}_1 n)]^{  1/(d+1) } }, e^{-x_0^2/(\tilde{c}_2\ell^d n)}\right) \right] \, , \label{CH_micro_canonical}
\end{align}
where $C_\Delta := 2 e^{\beta \Delta}  (e+3e\xi)  \left[ 2+( C_1 \Delta^{-1} \sqrt{n}) \right]$ with $C_1$ a constant that depends on $d$, $\alpha$, $\xi$ and $\ell_0$. } 
\end{prop}

\vspace*{0.5cm}
\noindent
Theorem~\ref{thm:main_theorem_Ensemble Equivalence}   is a direct result from this proposition. To show Theorem~\ref{thm:main_theorem_Ensemble Equivalence}  , we evaluate the upper bound of
\begin{align}
 | \langle \Omega \rangle_{{{E}},\Delta} - \langle \Omega \rangle_{\beta} |
  = \left| \int_{-\infty}^{\infty}\,  ( x - \langle \Omega \rangle_{\beta}  )   P_{{{E}},\Delta}(x) dx  \right|
\le  \int_{-\infty}^{\infty}\,  | x - \langle \Omega \rangle_{\beta}  |   P_{{{E}},\Delta}(x) dx  
  \, . \label{estimation_average_ensemble}
\end{align}
We apply Lemma~\ref{Prob:expectation} with $k=1$. From  Ineq.~\eqref{CH_micro_canonical}, we identify the parameter sets in Lemma~\ref{Prob:expectation} as follows:
\begin{align}
a=  \langle \Omega \rangle_{\beta} \, , \quad \{\gamma, \sigma, x_1\} = \{2/(d+1), (\tilde{c}_1n)^{\frac{1}{d+1}}, \log C_\Delta \} \quad {\rm and} \quad \{2, \tilde{c}_2\ell^d n, \log C_\Delta \}. \label{consider two parameter sets in Lemma}
\end{align}
We subsequently obtain  
\begin{align}
  | \langle \Omega \rangle_{{{E}},\Delta} - \langle \Omega \rangle_{\beta} | \le \max ( \tilde{\mathcal{B}}_1,\tilde{\mathcal{B}}_2  ) \, , 
\end{align}
where
\begin{align}
&\tilde{\mathcal{B}}_1:=\sqrt{\tilde{c}_1 n} (2\log C_\Delta)^{\frac{d+1}{2}}+2^{(d-1)/2} (d+1)\sqrt{\tilde{c}_1 n}\Gamma[(d+1)/2]\, ,    \notag \\
&\tilde{\mathcal{B}}_2:=\sqrt{2\tilde{c}_2\ell^d n \log C_\Delta}+\frac{1}{2}\sqrt{2\pi \tilde{c}_2\ell^d n }  \,  . \label{upp_bound_estimation_average_ensemble}
\end{align}
Here, we use $\Gamma(1/2)=\sqrt{\pi}$. Because $C_\Delta = \orderof{\sqrt{n}}$, the first terms in \eqref{upp_bound_estimation_average_ensemble} are dominant. 
We thus obtain the primary inequality in Theorem~\ref{thm:main_theorem_Ensemble Equivalence}  .
This completes the proof of Theorem~\ref{thm:main_theorem_Ensemble Equivalence}  . 

{~}\\

\subsection{Proof of Proposition~\ref{main_theorem_Ensemble Equivalence_clustering}.}
\noindent
Let $M$ be a positive even integer. Subsequently, we define the $M$th order moment of the observable
\begin{align}
  \mathcal{M}_{{{E}},\Delta}(M) := \int_{-\infty}^\infty (x - \langle \Omega \rangle_{\beta}  )^M P_{{{E}},\Delta}(x)  dx . \label{moment_function_micro_cano}
\end{align}
We note the following relation:
\begin{align}
\mathcal{M}_{{{E}},\Delta}(M) &= {1\over {\cal N}_{{{E}},\Delta}} \sum_{E_m \in ({{E}}- \Delta, {{E}} ] } [\,  \bra{E_m} ( \Omega - \langle \Omega \rangle_{\beta} ) \ket{E_m} \, ] ^M   \nonumber \\
              & \le {1\over {\cal N}_{{{E}},\Delta}}
                \sum_{E_m \in ({{E}}- \Delta, {{E}} ] } \bra{E_m} (\Omega - \langle \Omega \rangle_{\beta})^M \ket{E_m}  =  \left\langle  (\Omega - \langle \Omega \rangle_{\beta})^M \right\rangle_{{{E}},\Delta} \, ,
                \label{moment_dist_omega_MC}
\end{align}
where we used the following relation that is valid for any state $\ket{\psi}$ from the convexity of the function $x^M$ with even $M$
\begin{align}
\bra{\psi} ( \Omega - \langle \Omega \rangle_{\beta})  \ket{\psi}^M  \le \bra{\psi} (\Omega - \langle \Omega \rangle_{\beta})^M \ket{\psi} \, .
\end{align}
We next use the following relation for arbitrary non-negative operators $\tilde{O}$, which is discussed in Ref.~\cite{Tasaki2018} (We also provide the proof in the next subsection)
\begin{align}
  { \langle \tilde{O} \rangle_{{{E}},\Delta} \over  \langle \tilde{O} \rangle_{\beta}} \le 2 e^{\beta \Delta}  \left[ 2+ (C_1 \sqrt{n} / \Delta ) \right] \, . \label{mc_Can_difference_Z}
\end{align}
Applying the non-negative operator $\tilde{O}= (\Omega -\langle \Omega \rangle_{\beta})^M$ to this inequality, we obtain the following relation:
\begin{align}
  \mathcal{M}_{{{E}},\Delta}(M) \le \langle  (\Omega -\langle \Omega \rangle_{\beta})^M \rangle_{{{E}},\Delta}
  \le 2 e^{\beta \Delta}  \left[ 2+ (C_1 \sqrt{n} / \Delta) \right] \langle (\Omega -\langle \Omega \rangle_{\beta}  )^M \rangle_{\beta} \, .
\end{align}
By combining the inequality above with Markov's inequality, we obtain
\begin{align}
  P_{{{E}},\Delta}( | x - \langle \Omega \rangle_{\beta}| \ge x_0) & = P_{{{E}},\Delta}( ( x - \langle \Omega \rangle_{\beta})^M \ge x_0^M) \nonumber \\
 & \le { \mathcal{M}_{{{E}},\Delta}(M) \over  x_0^M } \le 2 e^{\beta \Delta}  \left[ 2+ (C_1 \sqrt{n} / \Delta) \right] { \langle (\Omega - \langle \Omega \rangle_{\beta})^M \rangle_{\beta} \over  x_0^M} \, .\label{P_tilde_Omega_bound}
\end{align}
From Lemma~\ref{lem:main_lemma_Anshu}, the moment $\langle (\Omega-\langle \Omega \rangle_{\beta})^M \rangle_{\beta}$ satisfies the inequality~\eqref{moment_bound_concentration}.
Hence, the inequality~\eqref{CH_micro_canonical} can be derived by choosing $M$ as in Eq.~\eqref{choice_of_m_para} such that inequality~\eqref{concentration_bound_clustering} holds.
This completes the proof of Proposition~\ref{main_theorem_Ensemble Equivalence_clustering}. $\square$

\subsection{Proof of the inequality~\eqref{mc_Can_difference_Z}} 
\noindent
We follow the proof in Ref.~\cite{Tasaki2018}. We start with the following inequality:
\begin{align}
\langle \tilde{O} \rangle_{{{E}},\Delta} &={1\over {\cal N}_{{{E}},\Delta}}  \sum_{E_m \in ({{E}}- \Delta , {{E}}] } \bra{E_m} \tilde{O} \ket{E_m} \notag \\
&\le {e^{\beta {{E}}} \over  {\cal N}_{{{E}},\Delta} } \sum_{E_m \in ({{E}}- \Delta , {{E}} ] }e^{-\beta E_m} \bra{E_m} \tilde{O} \ket{E_m}\notag \\
                                     &\le {Z_{\beta}  e^{\beta {{E}}} \over  {\cal N}_{{{E}},\Delta} } \sum_{E_m \in ( -\infty , \infty ) } { e^{-\beta E_m} \over Z_{\beta} }\bra{E_m} \tilde{O} \ket{E_m}
=
    (    Z_{\beta} e^{\beta {{E}}}/ {\cal N}_{{{E}},\Delta} )    \langle \tilde{O} \rangle_{\beta} \, .
\label{mc_Can_difference_Z2}
\end{align}

\noindent
To bound $(Z_{\beta} e^{\beta {{E}}} / {\cal N}_{{{E}},\Delta})$ from above, we consider that the concentration inequality~\eqref{concentration_bound_clustering} can be applied to the Hamiltonian $H$ because the Hamiltonian satisfies the criterion of $\Omega$, (\ref{Def:Omega_l_local}) by setting $\ell=\ell_0$. By applying $\Omega = H$ in the probability distribution (\ref{pdensity}), the concentration is bound for the distribution of the canonical distribution. This implies that the finite distribution is exponentially dominated by the regime around the average. Hence, the regime $\zeta_{\beta}$ exists that satisfies
  \begin{align}
  \tr \left[ \sum_{E_m \in \zeta_{\beta} } {e^{-\beta E_m} \over  Z_{\beta}}\ket{E_m} \bra{E_m} \right]  &\ge 1/2  \, , \\
    \zeta_{\beta} := (\, \langle H \rangle_{\beta} - C_1 \sqrt{n}/2 , & ~\langle H \rangle_{\beta} + C_1  \sqrt{n} /2\, ] \, , 
\end{align}
where $C_1$ depends only on $d$, $\xi$, and $\ell_0$. Let us define the following:
\begin{align}
  \tilde{Z}  := \tr \left[ \sum_{E_m \in \zeta_{\beta}  }e^{-\beta E_m} \ket{E_m} \bra{E_m} \right] \ge Z_{\beta}/2 \, . 
\end{align}
Subsequently, using a slightly extended regime $\zeta_{\beta}' := (\, \langle H \rangle_{\beta}-\Delta - C_1 \sqrt{n}/2\, ,  ~\langle H \rangle_{\beta} + \Delta + C_1  \sqrt{n} /2\, ]$, the quantity $ \tilde{Z}$ is bounded from above as follows:
\begin{align}
  \tilde{Z} &\le   \sum_{\nu\in \mathbb{Z}: \nu \Delta \in \zeta_{\beta}'   }{\cal N}_{\nu \Delta, \Delta} e^{-\beta \Delta (\nu-1) }.
  \label{upper_bound_tilde_Z_0}
\end{align}
From the definition of ${{E}}$~\eqref{Def:U_micro_en_re}, we have
\begin{align}
{\cal N}_{\nu \Delta, \Delta} e^{-\beta \Delta (\nu-1) } \le e^{\beta \Delta}  e^{-\beta {{E}}}{\cal N}_{{{E}} ,\Delta} ,
\end{align}
which reduces the inequality~\eqref{upper_bound_tilde_Z_0} to
\begin{align}
  \tilde{Z} &\le  e^{\beta \Delta}   \sum_{\nu\in \mathbb{Z}: \nu \Delta \in \zeta_{\beta}'   } e^{-\beta {{E}}}{\cal N}_{{{E}} ,\Delta} =   e^{\beta \Delta} \left[2+ (C_1 \sqrt{n} / \Delta )\right] e^{-\beta {{E}}}{\cal N}_{{{E}} ,\Delta} , \label{upper_bound_tilde_Z}
\end{align}
where we use the definition of ${{E}}$ in Eq.~\eqref{Def:U_micro_en_re}. We note $Z_{\beta}\le 2\tilde{Z}$ as well as the inequalities~\eqref{upper_bound_tilde_Z} and \eqref{mc_Can_difference_Z2} and subsequently arrive at the following relation:
\begin{align}
{\langle \tilde{O} \rangle_{{{E}},\Delta} \over  \langle \tilde{O} \rangle_{\beta}} \le  {2\tilde{Z}  e^{\beta {{E}}} \over  {\cal N}_{{{E}},\Delta}} \le 2 e^{\beta \Delta} \left[2+ (C_1\sqrt{n} / \Delta) \right] \, .
\end{align} 
This completes the proof. $\square$

\section{Weak ETH from Proposition~\ref{main_theorem_Ensemble Equivalence_clustering}}
\noindent
From the definition of \eqref{def:P_omega_x}, Proposition~\ref{main_theorem_Ensemble Equivalence_clustering} provides the probability such that a randomly chosen eigenstate $\ket{E_m}$ from the energy shell satisfies
\begin{align}
\frac{1}{n}\left| \bra{E_m}\Omega \ket{E_m}- \langle \Omega \rangle_{{{E}},\Delta} \right| \le n^{-\epsilon} \quad  (0<\epsilon <1/2). \label{ineq_weak_ETH}
\end{align}
Because this probability subexponentially converges to $1$ as $1-\exp \left [ -\mathcal{O}\left( n^{\frac{1-2\epsilon}{d+1}} \right)\right]$, we have $\bra{E_m}\Omega \ket{E_m}/n\simeq \langle \Omega \rangle_{{{E}},\Delta}/n$ with a probability of almost 1 for a sufficiently large $n$. 

\noindent
For a more quantitative discussion, we calculate the variance in the energy shell: 
\begin{align}
{\rm Var}_{{{E}},\Delta}(\Omega) := \frac{1}{{\cal N}_{{{E}},\Delta}} \sum_{E_m \in ({{E}}- \Delta, {{E}} ] }  (\bra{E_m} \Omega\ket{E_m})^2 -  \left( \frac{1}{{\cal N}_{{{E}},\Delta}} \sum_{E_m \in ({{E}}- \Delta, {{E}} ] }  \bra{E_m} \Omega\ket{E_m}\right)^2,
\end{align}
which is equivalent to $\mathcal{M}_{{{E}},\Delta}(2)- [\mathcal{M}_{{{E}},\Delta}(1)]^2$. 
Recall that the function $\mathcal{M}_{{{E}},\Delta}(M)$ has been defined in Eq.~\eqref{moment_function_micro_cano}.
Our task is to calculate 
\begin{align}
{\rm Var}_{{{E}},\Delta}(\Omega)= \mathcal{M}_{{{E}},\Delta}(2)- [\mathcal{M}_{{{E}},\Delta}(1)]^2  = \int_{-\infty}^\infty x^2 P_{{{E}},\Delta}(x)  dx - \left( \int_{-\infty}^\infty x P_{{{E}},\Delta}(x)  dx \right)^2 \le  \int_{-\infty}^\infty x^2 P_{{{E}},\Delta}(x)  dx,   \label{upp_bound_estimation_average_ensemble_weak_ETH00}
\end{align}
where we set $\langle \Omega \rangle_{\beta}=0$.
Because of the inequality~\eqref{CH_micro_canonical}, we can utilize Lemma~\ref{Prob:expectation} by choosing the parameters $(\gamma,\sigma,x_1)$ in Eq.~\eqref{consider two parameter sets in Lemma}. 
From the inequality~\eqref{ineq:Prob:expectation} with $k=2$, a straightforward calculation yields 
\begin{align}
\int_{-\infty}^\infty x^2 P_{{{E}},\Delta}(x)  dx  \le  \max ( \tilde{\mathcal{B}}'_1,\tilde{\mathcal{B}}'_2  )
\end{align}
with 
\begin{align}
&\tilde{\mathcal{B}}_1':=\tilde{c}_1 n (2\log C_\Delta)^{d+1}+2^{d+1}\tilde{c}_1 n  (d+1)! ,   \notag \\
&\tilde{\mathcal{B}}_2':=2\tilde{c}_2\ell^d n \log C_\Delta + 2 \tilde{c}_2\ell^d n    , \label{upp_bound_estimation_average_ensemble_weak_ETH}
\end{align}
where $C_\Delta=\orderof{\Delta^{-1}\sqrt{n}}$ was defined in Proposition~\ref{main_theorem_Ensemble Equivalence_clustering}.
By combining the inequalities~\eqref{upp_bound_estimation_average_ensemble_weak_ETH00} and \eqref{upp_bound_estimation_average_ensemble_weak_ETH},  we obtain
\begin{align}
{\rm Var}_{{{E}},\Delta}(\Omega/n) \le (1/n^2)  \max ( \tilde{\mathcal{B}}'_1,\tilde{\mathcal{B}}'_2  ) .
\end{align}
From the above inequality, we prove the inequality~(15) in the main manuscript.
Therefore, provided that $\Delta = 1/\poly(n)$, this estimation provides the upper bound of the variance of ${\rm Var}_{{{E}},\Delta}(\Omega/n)$ by $\orderof{\log^{d+1}(n)/n}$.

\section{Correlation length vs. inverse temperature in many-body localized systems}
\noindent
We herein discuss that in one-dimensional disordered systems, the clustering condition can break down at sufficiently low temperatures.
For such disordered systems, we consider many-body localized systems, where thermalization do not occur. 
Hence, we consider the $xy$ model with random magnetic fields:
\begin{align}
  H=\sum_{i=1}^{n-1} (3/4) \sigma_i^x   \sigma_{i+1}^x + (1/4) \sigma_i^y   \sigma_{i+1}^y +    \sum_{i=1}^n h_i \sigma_i^z  \, , \label{XY_model_H_random}
\end{align}
where each of $\{h_i\}_{i=1}^n$ is chosen randomly from a uniform distribution in $[-1,1]$. This system can be mapped onto a bilinear fermionic system; hence, the system cannot be regarded as a many-body localized system but as a system exhibiting the Anderson localization. However, we expect to extract the essential property even with this model. This model allows us to consider an large system by which an accurate correlation length can be computed. 

\begin{figure*}[]
\centering
\includegraphics[clip, scale=0.5]{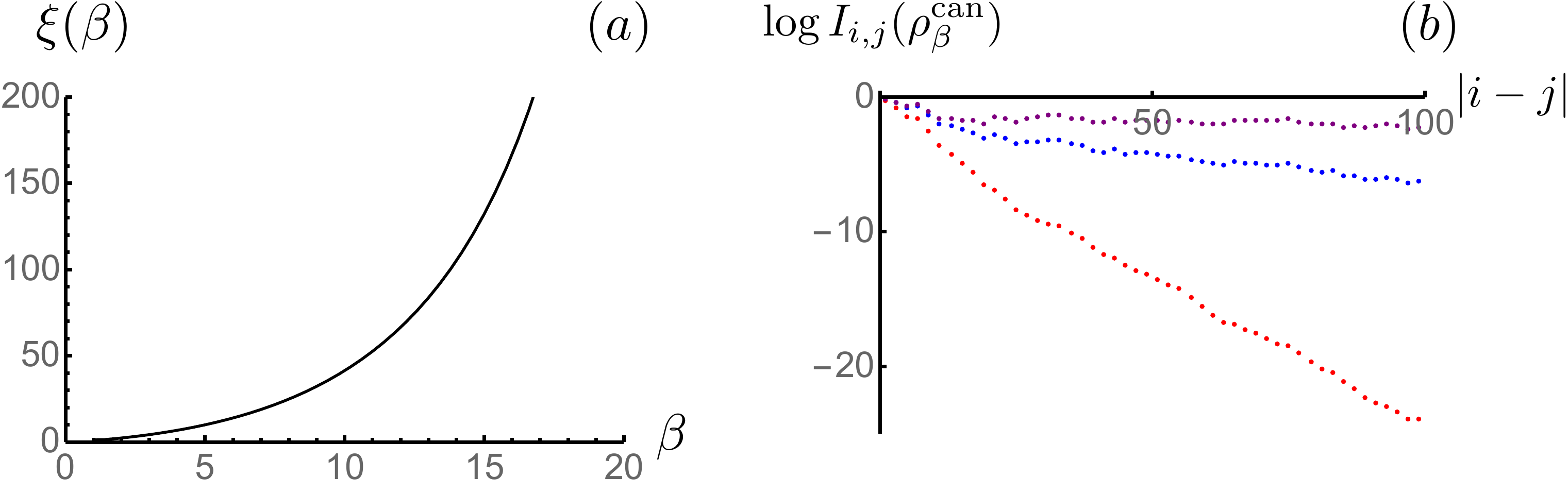}
\caption{(color online) Numerical demonstrations. The first figure (a) shows $\beta$-dependence of the correlation length $\xi(\beta)$ in the canonical state $\rho_{\beta}$ for one sample of the random Hamiltonian~\eqref{XY_model_H_random}. The correlation length $\xi(\beta)$ is calculated from the mutual information between the spin pairs of $\{r, n-r+1\}_{r=1}^{n/2}$ with $n=100$. 
The second figure (b) shows the distance dependence of the mutual information $I_{i,j}(\rho_\beta)$ for $\beta=3$ (red plots), $\beta=7$  (blue plots), $\beta=13$ (purple plots). 
}
\label{fig:MBL_corr}
\end{figure*}
\noindent
We consider the mutual information $I_{i,j}(\rho_\beta)$ between the sites $i$ and $j$ is defined as $I_{i,j}(\rho_\beta)$ for $1\le \beta \le 20$.
In Fig.~\ref{fig:MBL_corr} (a), we calculate the $\beta$-dependence of the correlation length $\xi(\beta)$.
Furthermore, we present the distance dependence of the mutual information $I_{i,j}(\rho_\beta)$ for $\beta=3$ (red plots), $\beta=7$  (blue plots), $\beta=13$ (purple plots) in Fig.~\ref{fig:MBL_corr} (b). 
From this plot, the correlation length diverges as $\beta$ increases and the clustering property breaks down at sufficiently low temperatures.

\section{Reduced density matrices of $\rho_{{{E}},\Delta}$ and $\rho_{\beta}$ and  Theorem~\ref{thm:main_theorem_Ensemble Equivalence}   in the main text}
\noindent
To compare the result from Theorem~\ref{thm:main_theorem_Ensemble Equivalence}   with the previous one by Br$\~a$ndao and Cramer~\cite{brandao2015equivalence},
we estimate the trace distance between the reduced density matrices of the canonical and micro-canonical states.
Let $\{B_s\}_{s=1}^{n_{\ell}}$ be the set of all $d$-dimensional $\ell\times \ell \times \cdots \times \ell$ hypercubes~(see Fig.~\ref{translation_invariance}), where we assume that the total number of hypercubes $n_{\ell}$ is $n/\ell^d$ (i.e., $n_{\ell}:=n/\ell^d$). We now denote the reduced density matrices within the hypercubes $\{B_s\}_{s=1}^{n_{\ell}}$ by $\{(\rho_{\beta})_{B_s}\}_{s=1}^{n_{\ell}}$ and $\{(\rho_{{{E}},\Delta})_{B_s}\}_{s=1}^{n_{\ell}}$. From theorem~2, we can prove 
\begin{align}
\frac{1}{n_{\ell}} \sum_{s=1}^{n_{\ell}}\left\| (\rho_{{{E}},\Delta})_{B_s}-(\rho_{\beta})_{B_s} \right\|_1  
\le&\frac{\max\left[c'_1 \ell^{-\frac{d}{2}} \log^{\frac{d+1}{2}} \bigl(\frac{\sqrt{n}}{\Delta}\bigr)  , c'_2\sqrt{ \log \bigl(\frac{\sqrt{n}}{\Delta}\bigr)}\right]  }{\sqrt{n_{\ell}}},
\label{thm1:op_ensemble_equiv_block}
\end{align}
where $\|\cdots\|_1$ denotes the trace norm (see below for the derivation). If we assume the translation invariance of the Hamiltonian \eqref{Hamiltonian_supple}, all the reduced density matrices $\{\rho_{\beta})_{B_s},(\rho_{{{E}},\Delta})_{B_s}\}_{l=1}^{n_{\ell}}$ do not depend on the index $s$. Hence, for an arbitrary hypercube $B$, the norm difference between $(\rho_{\beta})_{B}$ and $(\rho_{{{E}},\Delta})_{B}$ is smaller than $\orderof{1/\sqrt{n_{\ell}}}$. 
One of the open problem posed by Brand\~ao and Cramer is how large the block can be in order to guarantee the ensemble equivalence~\cite{brandao2015equivalence}. 
From the inequality~\eqref{thm1:op_ensemble_equiv_block}, provided that $\ell^d=\orderof{n^{1-\epsilon}}$ ($\epsilon>0$) and $\Delta = 1/\poly(n)$, the reduced density matrices between the canonical and the microcanonical ensembles are indistinguishable in the limit of $n\to \infty$, where the size dependence of the error behaves as $\orderof{1/\sqrt{n_{\ell}}}=\orderof{n^{-\epsilon/2}}$ with a logarithmic correction $\poly[\log(n)]$.

\noindent
The upper bound of \eqref{thm1:op_ensemble_equiv_block} qualitatively improves the results of previous studies~\cite{Muller2015,de2006quantum,Lima1972,lima1971equivalence,brandao2015equivalence,Tasaki2018}. In Refs~\cite{Muller2015,de2006quantum,Lima1972,lima1971equivalence}, the thermodynamic limit $n\to \infty$ has been considered, and the energy width $\Delta$ and block size $\ell$ are assumed to be independent of the system size $n$. In Ref~\cite{brandao2015equivalence}, the finite-size effect has been considered for the first time, where the LHS of \eqref{thm1:op_ensemble_equiv_block} has been bounded from above by $C \max(\sqrt{\ell^{d(d+1)}/n},n^{-1/(2d+2)})$ with $C=\poly[\log(n)]$ under the assumptions of the clustering and $\Delta = \orderof{\log^{2d} (n)}$. Subsequently, the block size $\ell^d$ can be as large as $\ell^d=n^{1/(d+1)-\epsilon}$ ($\epsilon>0$). In Ref.~\cite{Tasaki2018}, this upper bound for the LHS of \eqref{thm1:op_ensemble_equiv_block} has been improved to $\sqrt{\Delta^{-1}\ell^d/n^{1/2}}$ under some additional assumptions. Then, the ensemble equivalence holds for $\ell^d=\Delta\cdot n^{1/2-\epsilon}$ ($\epsilon>0$).

\begin{figure}[t]
\centering
{
\includegraphics[clip, scale=0.3]{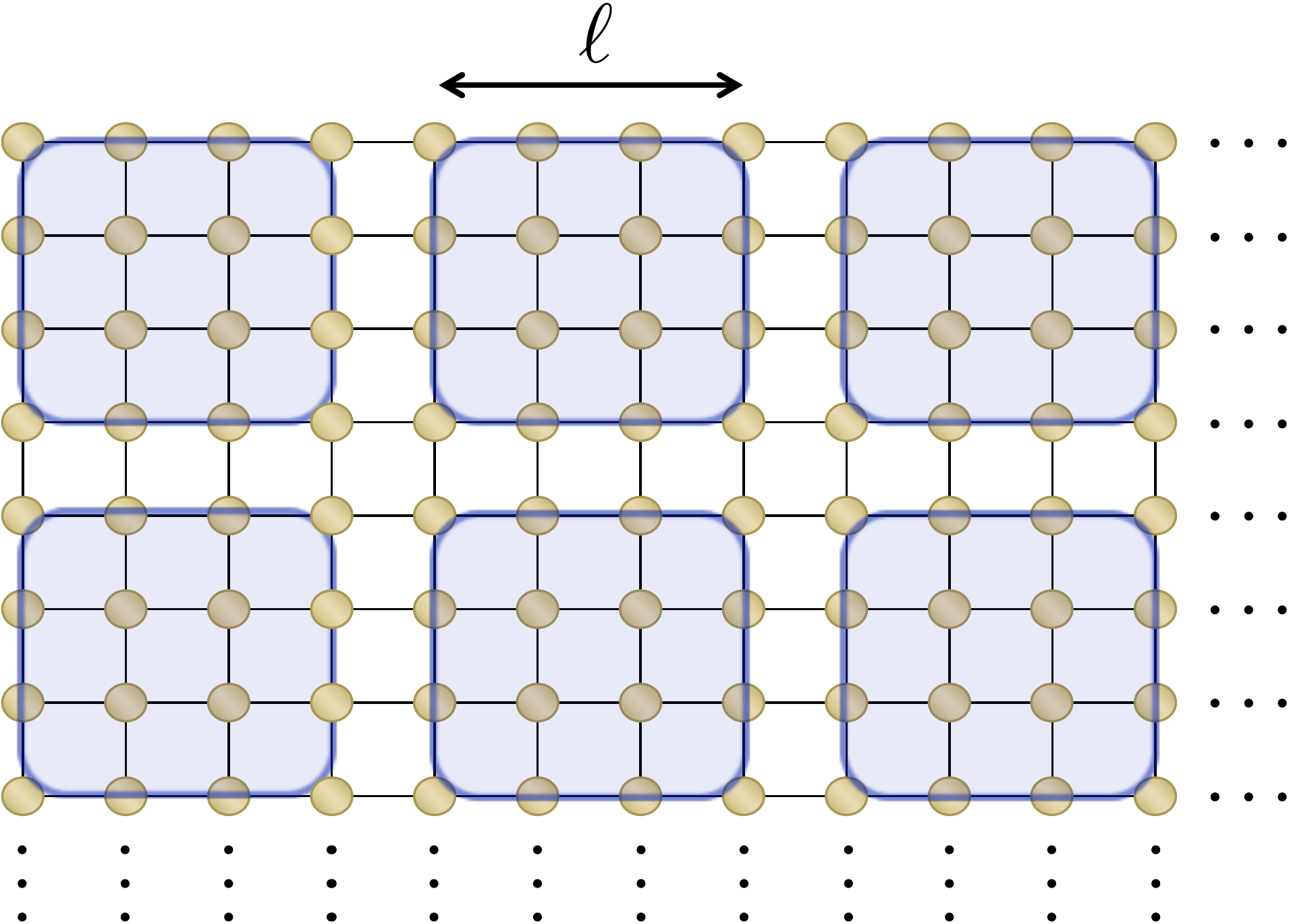}
}
\caption{(color online). Schematics of the setup in Ineq.~\eqref{thm1:op_ensemble_equiv_block} on a two-dimensional square lattice. We decompose the lattice to $n_{\ell}$ subsets $\{B_s\}_{s=1}^{n_{\ell}}$, each of which contains $\ell\times \ell$ spins ($\ell=4$ in the picture above). We focus on the reduced density matrices with respect to $\{B_s\}_{s=1}^{n_{\ell}}$, and estimate the norm difference in Ineq.~\eqref{thm1:op_ensemble_equiv_block} between the canonical and the microcanonical states. 
From the inequality~\eqref{thm1:op_ensemble_equiv_block}, as long as the block size is $\ell^d=\orderof{n^{1-\epsilon}}$ ($\epsilon>0$), the ensemble equivalence holds. 
}
\label{translation_invariance}
\end{figure}

\subsection{Proof of the inequality~\eqref{thm1:op_ensemble_equiv_block}} 
\noindent
We first consider that for an arbitrary hypercube $B_s$, there exists an operator $O_{B_s}$ with $\| O_{B_s}\|=1$ such that 
\begin{align}
\left\|(\rho_{{{E}},\Delta})_{B_s} - (\rho_{\beta})_{B_s} \right\|_1 = \tr \left[O_{B_s} \left( (\rho_{{{E}},\Delta})_{B_s} - (\rho_{\beta})_{B_s} \right)\right].
\end{align}
We subsequently choose $\Omega= \sum_{i=1}^n \Omega_i$ as 
\begin{align}
\Omega_i = O_{B_s} \for  i \in B_s,
\end{align}
which yields
\begin{align}
\frac{1}{n}\left|\langle \Omega \rangle_{{{E}},\Delta}   - \langle \Omega \rangle_{\beta} \right| 
=& \frac{\ell^d}{n}\sum_{s=1}^{n_{\ell}} \left|\langle O_{B_s} \rangle_{{{E}},\Delta}   - \langle O_{B_s} \rangle_{\beta} \right|  \notag \\
=& \frac{1}{n_{\ell}}\sum_{s=1}^{n_{\ell}} \left\|(\rho_{{{E}},\Delta})_{B_s} - (\rho_{\beta})_{B_s} \right\|_1, \label{block_norm_average}
\end{align}
where we use the assumption of $n_{\ell}=n/\ell^d$. 
By combining the inequality~\eqref{block_norm_average} with Theorem~\ref{thm:main_theorem_Ensemble Equivalence}  , we obtain the inequality~\eqref{thm1:op_ensemble_equiv_block}.

\newcommand{\Cl}{\mathcal{C}}

\section{Concentration bound from the clustering of correlation}  \label{Sec:Concentration bound from the clustering}
\noindent
We herein show the proof of Lemma~\ref{lem:main_lemma_Anshu} that reduces to the concentration bound~\eqref{concentration_bound_clustering}.
The proof that we address herein is essentially the same as that of Ref.~\cite{Anshu_2016} but is more general and simplified.
To account for the lattice geometry, we consider a generic graph $G=(V,E)$ with $|V|=n$, where each of the spin sits on the vertex.
For two arbitrary subsets $X,Y \subset V$, we define the distance $\dist(X,Y)$ as the minimum path length from $X$ to $Y$ on the graph.
For an arbitrary vertex $i \in V$, we define the set of $X_i^{(s)} \in V$ ($s\in \mathbb{N}$) as 
\begin{align}
X_i^{(s)} := \{j \in V | \dist(j,i) \le s\} \label{def:X_i^s}.
\end{align}
Each of the local terms $\{\Omega_i\}_{i\in V}$ in Eq.~\eqref{Def:Omega_l_local} is supported on the subset $X_i^{(\ell)}$. 
We introduce a geometric parameter $\alpha$ that depends on the lattice structure as 
\begin{align}
|X_i^{(s)}| \le  \alpha s^d, \label{def:X_i^s_alpha}
\end{align}
where $d$ is the spatial dimension of the lattice.

\subsection{Proof of Lemma~\ref{lem:main_lemma_Anshu}}
\noindent
We define $\delta \Omega_i :=\Omega_i -\langle \Omega_i \rangle_\rho$ for $\forall i \in V$ that yields $\langle \delta \Omega_i \rangle_\rho=0$.
We calculate the upper bound of $\langle (\Omega - \langle \Omega \rangle_\rho)^M \rangle_\rho $, where $M$ is a positive even integer.
Next, we decompose 
 \begin{align}
   \langle (\Omega - \langle \Omega \rangle_\rho )^M \rangle_\rho = \sum_{i_1,i_2,\ldots i_M \in V} \langle \delta\Omega_{i_1} \, \delta \Omega_{i_2} \, \cdots \,\delta\Omega_{i_M} \rangle_\rho .
\end{align}
We subsequently define $\ell_{i_1,\ldots,i_M}$ as 
 \begin{align}
   \ell_{i_1,\ldots,i_M}   := \max_{1\le q \le M} \left[ \dist (i_q, \{i_r\}_{r\neq q} )\right].
\end{align} 
That is, the most spatially isolated vertex in $\{i_1,\ldots,i_M\}$ is separated from the other vertices by a distance $\ell_{i_1,\ldots,i_M}$.
If $\ell_{i_1,\ldots,i_M}=\tilde{l}$, there exists $i_q \in \{i_1,\ldots,i_M\}$ such that $\dist (i_q, \{i_r\}_{r\neq q}) = \tilde{l}$. 
Because each of the $\delta\Omega_{i}$ is supported on the subset $X_i^{(\ell)}$ from the assumption (\ref{Def:Omega_l_local}), the two operators $\delta\Omega_{i_q}$ and $\delta\Omega_{i_1}\cdots \delta\Omega_{i_{q-1}} \delta\Omega_{i_{q+1}} \cdots \delta\Omega_{i_M} $ are separated at the least by the distance $\tilde{l}-2\ell$.
Hence, for $\tilde{l} \ge 2\ell+r$, the $(r,\xi)$-clustering condition yields
 \begin{align}
 &   | \langle \delta \Omega_{i_1}\, \delta\Omega_{i_2}\,  \cdots \, \delta\Omega_{i_M} \rangle_\rho - \langle \delta\Omega_{i_q} \rangle_\rho  \langle \delta\Omega_{i_1}\, \cdots \, \delta\Omega_{i_{q-1}}\, \delta \Omega_{i_{q+1}} \, \cdots \, \delta\Omega_{i_M}  \rangle_\rho| \notag \\
    =&   | \langle \delta \Omega_{i_1}\, \delta\Omega_{i_2}\,  \cdots \, \delta\Omega_{i_M} \rangle_\rho |
   \le 2^M e^{-(\tilde{l}- 2\ell)/\xi},
\label{clustering_inequality_1}
\end{align}
where we use $\|\delta \Omega_i\|\le 2$ from $\|\Omega_i\|\le1$ for $\forall i \in V$. 

\vspace*{0.5cm}
\noindent
We define $\Cl_{\tilde{l}}$ ($\Cl_{\le \tilde{l}}$) as the set of string $(i_1,\ldots,i_M)$ such that $\ell_{i_1,\ldots,i_M}= \tilde{l}$ ($\ell_{i_1,\ldots,i_M}\le \tilde{l}$). 
Here, we take the element order in the string into account. For instance, we count $(i_1, i_2)$ and $(i_2, i_1)$ as a different string if $i_1\neq i_2$. We subsequently bound $\langle (\Omega - \langle \Omega \rangle_\rho)^M \rangle_\rho$ from above as follows: 
 \begin{align}
   \langle (\Omega - \langle \Omega \rangle_\rho)^M \rangle_\rho &= \sum_{\tilde{l}=0}^\infty  \sum_{(i_1,i_2,\ldots i_M)\in \Cl_{\tilde{l}}} \langle \delta\Omega_{i_1}\, \delta\Omega_{i_2} \, \cdots \, \delta\Omega_{i_M}  \rangle_\rho \notag \\
                                                                 &= \sum_{\tilde{l}=0}^{2\ell+r-1}  \sum_{(i_1,i_2,\ldots i_M)\in \Cl_{\tilde{l}}} \langle \delta\Omega_{i_1}\, \delta\Omega_{i_2} \, \cdots \, \delta\Omega_{i_M}  \rangle_\rho   +  \sum_{\tilde{l}=2\ell+r}^{\infty}  \sum_{(i_1,i_2,\ldots i_M)\in \Cl_{\tilde{l}}}
\langle \delta\Omega_{i_1}\, \delta\Omega_{i_2} \, \cdots \, \delta\Omega_{i_M}  \rangle_\rho  \notag \\                                                               
                                                              &\le 2^M \sum_{\tilde{l}=0}^{2\ell+r-1}  \sum_{(i_1,i_2,\ldots i_M)\in \Cl_{\tilde{l}}} 1 +  2^M\sum_{\tilde{l}=2\ell+r}^{\infty}  \sum_{(i_1,i_2,\ldots i_M)\in \Cl_{\tilde{l}}}e^{-(\tilde{l}- 2\ell)/\xi} \notag \\
&\le 2^M \sum_{\tilde{l}=0}^{3\ell}  \sum_{(i_1,i_2,\ldots i_M)\in \Cl_{\tilde{l}}} 1 +  2^M\sum_{\tilde{l}=3\ell+1}^{\infty}  \sum_{(i_1,i_2,\ldots i_M)\in \Cl_{\tilde{l}}}e^{-(\tilde{l}- 2\ell)/\xi}   \notag \\
                                                                 &\le 2^M \sum_{(i_1,i_2,\ldots i_M)\in \Cl_{\le 3\ell}} 1 +  2^M\sum_{\tilde{l}=3\ell+1}^{\infty}  \sum_{(i_1,i_2,\ldots i_M)\in \Cl_{\tilde{l}}}e^{-\tilde{l}/(3\xi)} \notag \\
                                                                 &\le 2^M \sum_{(i_1,i_2,\ldots i_M)\in \Cl_{\le 3\ell}} 1 +  2^M\sum_{\tilde{l}=3\ell+1}^{\infty}  \sum_{(i_1,i_2,\ldots i_M)\in \Cl_{\le \tilde{l}}                                          }e^{-\tilde{l}/(3\xi)}
                                                                   \, . \label{mth_moment_H}
\end{align}
Herein, we use the inequality~\eqref{clustering_inequality_1} in the first inequality as well as the relation $\|\delta \Omega_i\|\le 2$. Furthermore, in the second last inequality, we use $e^{-(\tilde{l}-2\ell)/\xi} \le e^{-\tilde{l}/(3\xi)}$ for $\tilde{l}\ge 3\ell$. At the last inequality, we use a trivial relation where $\Cl_{\le \tilde{l}}$ is a larger set than $\Cl_{\tilde{l}}$. 

\begin{figure}[t]
\centering
{
\includegraphics[clip, scale=0.5]{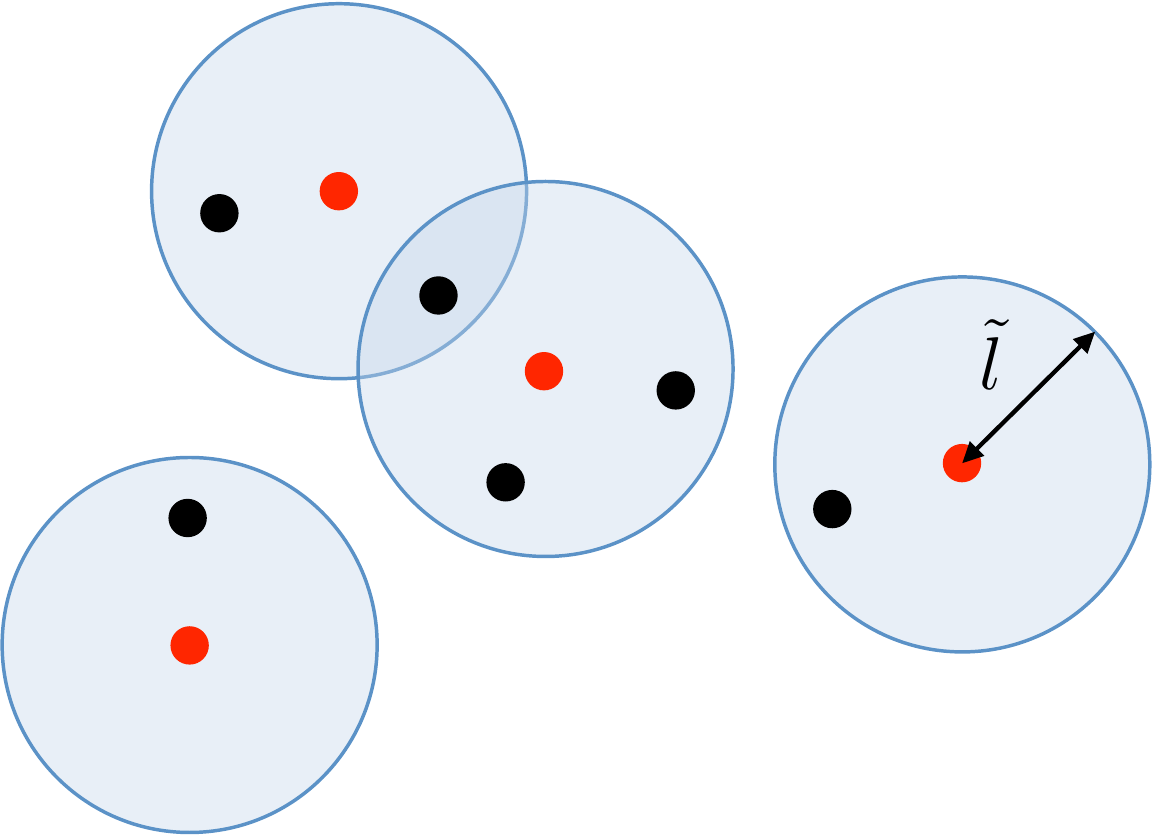}
}
\caption{(color online). Schematics of the core vertices (red dots) and the surrounding vertices (black dots). 
Around each of the core vertices within the distance $\tilde{l}$ (blue shaded region), there exists at least one surrounding vertex. 
Any string $\{i_1,\ldots,i_M\}\in \Cl_{\le \tilde{l}}$ can be decomposed into $u$ core vertices and $(M-u)$ surrounding vertices with $u\le M/2$.
}
\label{Fig:core_vertex}
\end{figure}

\vspace*{0.5cm}
\noindent
To count the total number of strings in the set $\Cl_{\le \tilde{l}}$, we follow the strategy below.
We first choose $u$ ``core'' vertices $(i_{q_1}, \ldots,i_{q_u})$ from $n$ vertices. We next assign the other $(M-u)$ vertices. 
We refer to the $(M-u)$ vertices $\{i_1,\ldots,i_M\} \setminus \{i_{q_1}, \ldots,i_{q_u}\}$ as the ``surrounding'' vertices.
Each of the core vertices contains at least one surrounding vertex within the distance $\tilde{l}$ around it (otherwise, the distance $\ell_{i_1,\ldots,i_M}$
exceeds $\tilde{l}$. See the schematics in Fig.~\ref{Fig:core_vertex}). 
Mathematically, for a core vertex $i_{q_1}$, there exists $i_r \in \{i_1,\ldots,i_M\} \setminus \{i_{q_1}, \ldots,i_{q_u}\}$ such that 
 \begin{align}
\dist (i_{q_1}, i_r) \le \tilde{l} \quad {\rm or} \quad i_r \in X^{(\tilde{l})}_{i_{q_1}}
\end{align}
with $X^{(\tilde{l})}_{i_{q_1}}$ defined in Eq.~\eqref{def:X_i^s}.
This constraint implies that the number of core vertices is smaller than $M/2$ (i.e., $u\le M/2$). 
We note that any string in $\Cl_{\le \tilde{l}}$ can be described by the formalism above.
Thus, our task is to count all the possible arrangements of 1) the core vertices, 2) the surrounding vertices for $u=1,2,\ldots, M/2$, and 3) the order of elements in the string.
For a fixed $u$, it is bounded from above as follows:
\begin{enumerate}
\item{} The number of possible locations of the core vertices is clearly smaller than $n^u$. 
\item{} After the locations of the core vertices are determined, there are at the most $|X^{(\tilde{l})}_{i_{q_1}} \cup X^{(\tilde{l})}_{i_{q_2}} \cup \cdots \cup X^{(\tilde{l})}_{i_{q_u}}| \le u (\alpha \tilde{l}^d)$ methods of positioning, at which each of the surrounding vertices can be placed. In total, the number of possible arrangements of the surrounding vertices is smaller than $[u (\alpha \tilde{l}^d)]^{M-u}$.
\item{} We next consider the string order. Inside each set of the core and surrounding vertices, the order is already considered by the two estimations above. Hence, we must consider the order of the types of vertices (i.e., ``core'' and ``surrounding''). The number of this combination is $\binom{M}{u}$. 
\end{enumerate}
We thus obtain
 \begin{align}
\sum_{(i_1,i_2,\ldots i_M)\in \Cl_{\le \tilde{l}}} 1\le \sum_{u=1}^{M/2} \binom{M}{u} n^u [u (\alpha \tilde{l}^d)]^{M-u}
&\le  [M n (\alpha \tilde{l}^d)/2]^{M/2} \sum_{u=1}^{M/2} \binom{M}{u}  \notag \\
&\le  2^M [M n (\alpha \tilde{l}^d )/2]^{M/2} =  [2 M n (\alpha \tilde{l}^d)]^{M/2}  .\label{counting_cluster_Cl}
\end{align}
\noindent 
By combining the inequality~\eqref{counting_cluster_Cl} with \eqref{mth_moment_H}, we obtain
 \begin{align}
\langle  [\Omega]^M \rangle_\rho &\le [8 M n \alpha (3\ell)^d]^{M/2}  +  \sum_{\tilde{l}=3\ell+1}^{\infty}  [8 M n (\alpha \tilde{l}^d)]^{M/2}e^{-\tilde{l}/(3\xi)}    \notag \\
&\le  [8\alpha M n   (3\ell)^d]^{M/2}  +  3\xi [8\alpha  n M^{d+1} (3\xi d /2)^d]^{M/2},
\end{align}
where the second inequality is given from 
 \begin{align}
 \sum_{\tilde{l}=3\ell+1}^{\infty}  [8 M n (\alpha \tilde{l}^d)]^{M/2}  e^{-\tilde{l}/(3\xi)} 
  &\le  (8\alpha M n)^{M/2} \int_0^\infty x^{dM/2}e^{-x/(3\xi)}dx \notag \\
  &= (8\alpha M n)^{M/2}  (3\xi)^{dM/2+1} (dM/2)!\le 3\xi [8\alpha  n M^{d+1} (3\xi d /2)^d]^{M/2}  .
  \end{align}
Note that we have $s! \le s^s$ for an arbitrary integer $s$.
This completes the proof of Lemma~\ref{lem:main_lemma_Anshu}. $\square$


\end{widetext}

\end{document}